\documentclass[a4paper,12pt]{article}

\usepackage{graphicx}
\usepackage{epstopdf}

\usepackage{amssymb,amsmath}
\usepackage{appendix}
\usepackage{wrapfig}

\usepackage{subcaption}
\usepackage{datetime}
\usepackage{natbib}
\allowdisplaybreaks

\newcommand{\bb}{\boldsymbol}

\newcommand{\deriv}[2]{\frac{\text{d}#1}{\text{d}#2}}

\newcommand{\Dtau}{\Delta \tau}

\newcommand{\wi}{.48}

\begin{document}

\title{A symplectic integrator for the symmetry reduced and regularised planar 3--body problem with vanishing angular momentum}
\date{\today}
\author{Danya Rose \and Holger R. Dullin \\ School of Mathematics and Statistics, The University of Sydney}

\maketitle

\begin{abstract}
We construct an explicit reversible symplectic integrator for the 
planar 3-body problem
with zero angular momentum. 
We start with a Hamiltonian of the planar 3-body problem that is globally regularised and fully symmetry reduced.
This Hamiltonian is a sum of 10 polynomials each of which can be integrated exactly,
and hence a symplectic integrator is constructed.
The performance of the integrator is examined with three numerical examples:
The figure eight, the pythagorean orbit, and a periodic collision orbit.

\end{abstract}

\section{Introduction}

It is well known that the flow $\Psi_H^t$ of a  Hamiltonian $H$ of the form 
$H = T(p) + V(q)$ with conjugate variables $q$ and $p$
can be approximated by splitting it into the integrable flow $\Psi_T^t$ of $T(p)$ 
and the integrable flow $\Psi_V^t$ of $V(q)$ and observing that
$\Psi_H^t = \Psi_T^t \circ \Psi_V^t + O(t^2)$, see, e.g.~ \cite{ChannellNeri1996,HaLuWa02,mclachlan2002,LeimkuhlerReich04}.
Thus a first order explicit symplectic integrator
is obtained, and higher order methods can be constructed along similar lines \cite{Yoshida1990}.
In an integrable and separable Hamiltonian system of the form $H = H_1(q_1, p_1) + H_2(q_2, p_2)$ such splitting 
gives the exact identity $\Psi_H^t = \Psi_{H_1}^t \circ \Psi_{H_2}^t $.
If instead the Hamiltonian is a product $H = H_1(q_1, p_1) H_2(q_2, p_2)$ again the system is
integrable with integrals $H_1$ and $H_2$ and the flow can be written as
\[
	\Psi_H^t = \Psi_{H_1}^{t H_2} \circ \Psi_{H_2}^{t H_1} =  \Psi_{H_2}^{t H_1} \circ  \Psi_{H_1}^{t H_2}  \,.
\]
In the superscript $tH_i$ denotes multiplication of $t$ by the (constant) value of $H_i$.
A monomial Hamiltonian is a special case that has the same structure. There are obvious 
generalisations to more degrees of freedom.
Hence  any polynomial Hamiltonian is a sum of integrable monomial Hamiltonians, 
and thus a splitting integrator can be constructed. 
Symplectic integration of polynomial Hamiltonians has been discussed in 
\cite{shi93,gjaja94,ChannellNeri1996,Blanes2002,quispel2004}.
{ In a recent paper by \cite{blanesIserles2012} various methods for time step control 
in geometric integrators are constructed and discussed. These methods
could be useful in order to implement variable time stepping on top of our method, 
see the discussion in section \ref{subsec:timestep}.
} 

In this paper we apply these methods to the polynomial Hamiltonian of the 
globally regularised and symmetry reduced 3-body problem at angular momentum zero.
For a review of numerical and regularisation methods in the $n$-body problem we refer to 
\cite{Tanikawa2007}.
It is well known that binary collisions in the 3-body problem can be regularised.
Regularisation consists of a canonical transformations which essentially extracts a square
root near collision, and of a scaling of time so that the approach to the collision is slowed down.
The classical simultaneous regularisation of the (spatial) 3-body problem is due to 
\cite{Heggie74}.
This increases the dimension of phase space from 18 to 24. 
Instead we would like to decrease the dimension of phase space by using reduction at the same time 
as regularisation. The simultaneous regularisation of the planar 3-body problem is due to 
\cite{lemaitre64},
and we use a version due to 
\cite{Waldvogel1982}. This is a symmetric simultaneous regularisation 
of the symmetry reduced planar 3-body problem and has the smallest possible 6-dimensional phase space.
The resulting Hamiltonian is a polynomial of up to degree 6 in the canonical variables.
A modern extension of these regularising transformations has recently been given 
by 
\cite{MoMo12}, however, their Hamiltonians are not polynomial but rational.

Our paper applies the methods for construction of an explicit symplectic integrator 
to Waldvogel's Hamiltonian with angular momentum zero. 
We also describe how a similar integrator could be constructed for Heggie's Hamiltonian, which 
works for non-zero angular momentum and in the spatial problem.

\section{The 3-body Hamiltonian}

\label{sec:coordinates}
The classical 3-body problem has long been studied, but still many open questions regarding its dynamics remain. 
For many questions, e.g.\ the study of relative periodic orbits, it is useful to  reduce by translational and rotational symmetries, so that the absolute rotation of an orbit can be separated from shape dynamics in the centre of mass frame.
Moreover, to study collision or near-collision orbits it is essential to perform (global) regularisation of the binary collisions.
Following 
 \cite{Waldvogel1982} we are going to do both.

\begin{figure}
	\centering
	\includegraphics[width=8cm]{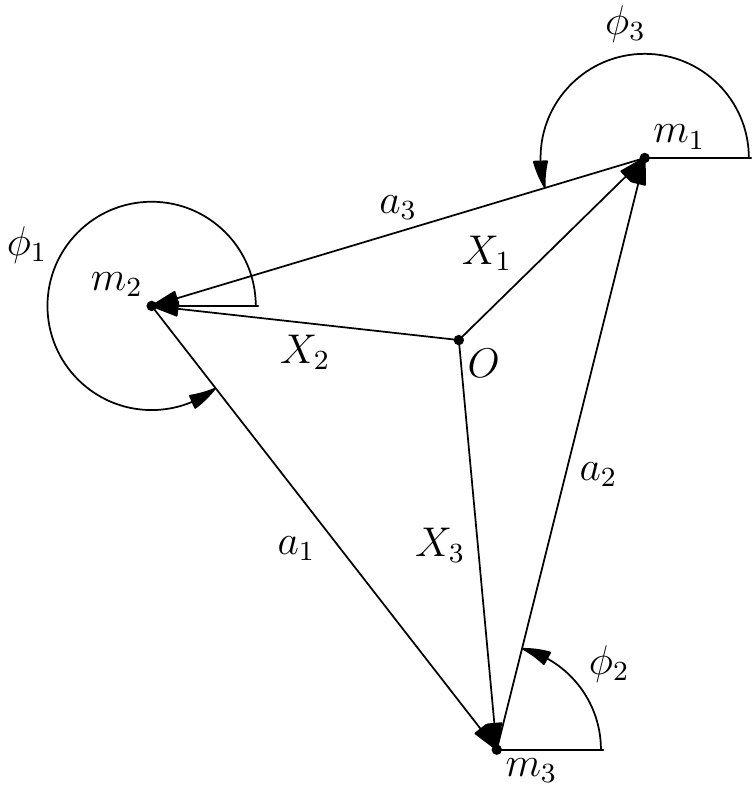}
	\caption{Coordinates of a triangle in the plane with centre of mass at the origin $O$.}
	\label{fig:1sym-reduce}
\end{figure}

If the position and momentum of mass $m_j$, for $j = 1,2,3$, are given by complex Cartesian coordinates $X_j$ and $P_j$ respectively, we can transform into symmetry-reduced coordinates such that $$X_l - X_k = a_j e^{\phi_j},$$ where $a_j = |X_l - X_k|$ is the length of the triangle's side opposite to $m_j$, $\phi_j$ is the angle of that side in the original coordinate system (in the direction of $m_k$ to $m_l$), as illustrated in figure \ref{fig:1sym-reduce}, and $(j,k,l)$ represents cyclic permutations of $(1,2,3)$.

This reduction results in coordinates $a_j$ and $\phi = \frac{1}{3} (\phi_1+\phi_2+\phi_3)$, which represents the orientation angle of the triangle with respect to the original choice of Cartesian coordinates, and corresponding canonical momenta $p_j$ and $p_\phi$. The Hamiltonian rewritten in these coordinates is independent of $\phi$, so $p_\phi$ is a constant of motion. Hamilton's equations for $(a_j,\phi,p_j,p_\phi)$ give the reduced dynamics, including a differential equation for $\phi$ which may be integrated along to 
be able to recover the unreduced position of the triangle.

\begin{figure}
	\centering
	\includegraphics[width=6cm]{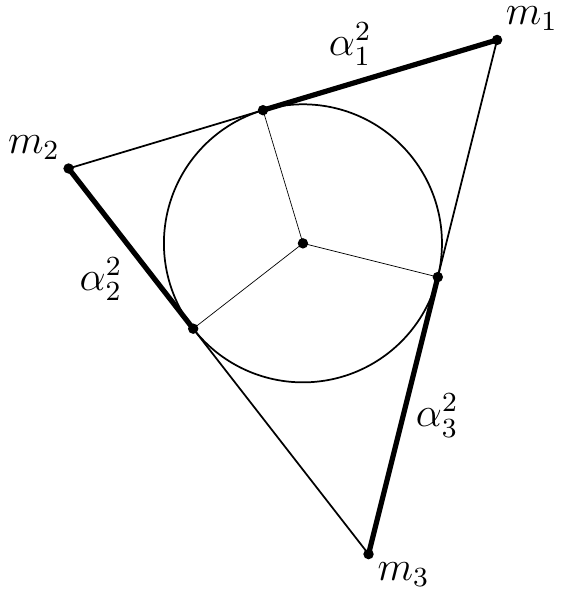}
	\caption{Physical significance of the regularised coordinates $\alpha_j$.}
	\label{fig:2reg}
\end{figure}

The globally regularising transformation, illustrated in figure \ref{fig:2reg}, goes from symmetry-reduced to regularised coordinates, simultaneously regularising all the binary collisions. Define $\alpha_j$ for $j = 1,2,3$ such that $a_j = \alpha_k^2 + \alpha_l^2$. In this way $\alpha_j^2$ is the distance from $m_j$ to the point where the incircle of the triangle touches the sides adjacent to $m_j$.

The space of coordinates $(a_j)$ is the space of all triangles, not accounting for orientation. Orientation is taken to be positive if, going clockwise around the triangle, the masses are encountered in a cyclic permutation of $(1,2,3)$ or negative otherwise. The space of all possible \emph{oriented} triangles is called the shape space, and the space of $(\alpha_j)$ is a four-fold covering of this space, in which the sign of the product $\alpha_1 \alpha_2 \alpha_3$ determines the orientation of the triangle. Thus the triangle formed by $(\alpha_1,\alpha_2,\alpha_3)$ is the same as the ones formed by $(\alpha_1,-\alpha_2,-\alpha_3)$, $(-\alpha_1,\alpha_2,-\alpha_3)$ and $(-\alpha_1,-\alpha_2,\alpha_3)$.

Canonically conjugate momenta $\pi_j$ are introduced using a generating function.
Finally the time scaling
\begin{equation}
	\label{eqn:timescaling}
	\deriv{t}{\tau} = a_1 a_2 a_3
\end{equation}
together with Poincar\'e's trick to make this Hamiltonian yields the regularised and symmetry reduced polynomial Hamiltonian
$$K = (H-h)a_1 a_2 a_3,$$ where $H$ is the original Hamiltonian written in the new coordinates and $h = H(\bb{\alpha}_0,\bb{\pi}_0)$ is the energy corresponding to the initial conditions $(\bb{\alpha}_0,\bb{\pi}_0)$, so only those solutions for which $K \equiv 0$ are physically meaningful.

The Hamiltonian of the zero-angular momentum 3-body problem in regularised coordinates is
\begin{equation}
	\label{eqn:K}
	K = K_0 - h a_1 a_2 a_3,
\end{equation}
where
\begin{equation}
	\label{eqn:K0}
	K_0 = \frac{1}{8} \bb{\pi}^T B(\bb{\alpha}) \bb{\pi} - \sum m_k m_l a_k a_l,
\end{equation}
in which
\begin{align*}
	\bb{\pi} &=
	\begin{pmatrix}
		\pi_1 &
		\pi_2 &
		\pi_3
	\end{pmatrix}^T \\
	\bb{\alpha} &=
	\begin{pmatrix}
		\alpha_1 &
		\alpha_2 &
		\alpha_3
	\end{pmatrix}^T \\
	B(\bb{\alpha}) &=
	\begin{pmatrix}
		A_1 & B_3 & B_2 \\
		B_3 & A_2 & B_1 \\
		B_2 & B_1 & A_3
	\end{pmatrix},
\end{align*}
where
\begin{align*}
	A_j &= \dfrac{a_j}{m_j}\alpha^2+\dfrac{a_k}{m_k}\alpha_l^2+\dfrac{a_l}{m_l}\alpha_k^2, \\
	B_j &= -\dfrac{a_j}{m_j}\alpha_k\alpha_l \text{\ \ and} \\
	\alpha^2 &= \alpha_1^2 + \alpha_2^2 + \alpha_3^2.
\end{align*}
{ 
The sum in \eqref{eqn:K0} (and any hereafter where the index of summation is unspecified) is over cyclic permutations of $(1,2,3)$, so that $(j,k,l)$ is replaced by $(1,2,3)$, $(2,3,1)$, and $(3,1,2)$ in turn, and then the three corresponding terms are added together. When there is no summation the indices $(j,k,l)$ take on the three possible cyclic
permutations in turn, as, e.g., in the definition of $A_j$ and $B_k$ above.
} 

The new Hamiltonian is a polynomial in $\bb{\alpha}$ and $\bb{\pi}$, and thus Hamilton's equations of motion for this system can be integrated with an explicit symplectic integrator obtained by splitting into monomials. As we are going to show in the next section it is more efficient to split into certain polynomials whose flow can be exactly solved.

\section{Construction of the Symplectic Integrator}

\label{sec:integrator}

The basic building blocks of the integrator are the exact solutions for monomial Hamiltonians $H_{mn} = q^m p^n$
in one degree of freedom.
The flow of this Hamiltonian for $m \ne n$ is
\begin{equation}
	\label{eqn:polysol1}
	\psi_{mn}^t(q,p) = ( q \beta^n, p \beta^{-m}), \quad \mbox{ where } \quad \beta = \left(1 + (n-m) q^{m-1} p^{n-1} t\right)^{\frac{1}{n-m}}
\end{equation}
while for $m = n$ it is
\begin{equation}
	\label{eqn:polysol2}
	\psi_{m}^t(q,p) = ( q \beta, p / \beta), \quad \mbox{ where } \quad \beta = \exp( m  (qp)^{m-1} t )\,.
\end{equation}
For the Hamiltonian we are studying the cases that occur are $n=m=1$, $n=m=2$, and $m=3$, $n=1$.
We also recall that if the Hamiltonian is a function of positions or momenta only 
(with any number of degrees of freedom) the flows are
\begin{equation}
\label{eqn:porqonly}
   \psi_{T(p)}^t(q, p) = ( q + (\nabla_p T) t, p),
   	 \quad \mbox{ and } \quad
   \psi_{U(q)}^t(q,p) = (q, p - (\nabla_q U) t) \,.
\end{equation}

\subsection{Integrable Polynomial Hamiltonians}
\label{subsec:monomialintegration}
The basic building blocks just mentioned are now combined to form integrators for the terms that appear in the Hamiltonian $K$.
The main observation is that if the Hamiltonian is a product of 
factors that depend on disjoint groups of degrees of freedom, then each factor is a constant of motion.
Each of the factors in our case is either 
depending on momenta or positions only (denoted by $T(p)$ or $U(q)$)
or it is a single monomial in one degree of freedom (denoted by $H_{mn})$
or a sum of monomials of disjoint degrees of freedom (denoted by $G$).

We now list the cases that are relevant in our case (recall that each of the factors depends on disjoint 
groups of degrees of freedom):
\begin{subequations}
\begin{alignat}{2}
\label{eqn:Ha}
& H_a  = T H_{nm} , \quad && \psi_a  = \psi_T^{t H_{nm}} \circ \psi_{nm}^{t T } \\
\label{eqn:Hb}
& H_b  = T V , \quad            && \psi_b  = \psi_T^{t V} \circ \psi_V^{t T } \\
\label{eqn:Hc}
& H_c  = G H_{nm}, \quad  && \psi_ c  = \psi_{nm}^{t G} \circ  \psi_G^{t H_{nm}}
\end{alignat}
\end{subequations}
where $G$ is a Hamiltonian which is the sum of Hamiltonians depending on disjoint degrees of freedom
$G = H_1(q_1, p_1) + H_2(q_2, p_2)$ and thus $\psi_G = \psi_{H_1} \circ \psi_{H_2}$.
Note that all these formulas are exact, and that the order of composition is irrelevant since the flows commute
and the individual factors are constants of motion.

\subsection{Splitting}
\label{subsec:splitting}
Let us now explain how to split $K$ \eqref{eqn:K} into such terms. It is a polynomial Hamiltonian of degree $6$ in $\bb{\alpha}$ and $\bb{\pi}$ with 34 monomials. There are $13$ monomials, dependent only on $\bb{\alpha}$, of degrees $6$ and $4$, which may be treated as a single stage. The remaining $21$ terms may be grouped such that only $9$ more stages are necessary to approximate the flow of the full Hamiltonian to first order in the time step in $10$ stages. Let $K = \sum_{i=0}^{9}H_i$, where we set $M_j = m_k m_l$ and $N_j = \frac{1}{m_k} + \frac{1}{m_l}$. Then the splitting is
\begin{equation}
	\label{eqn:splitting}
	\begin{array}{rlrl}
		H_0 =& - \sum M_j \alpha_j^4 - \left( \sum M_j \right) \left( \sum \alpha_k^2 \alpha_l^2 \right) - h a_1 a_2 a_3 &=& \ \ \ \ \        C_{0} \\
		H_1 =& \frac{1}{8} \left(N_2 \alpha_2^2 + N_3 \alpha_3^2\right) \alpha_1^2 \pi_1^2                               &=& \ \ \frac{1}{8}\ C_{1,23}\ C_{1,1} \\
		H_2 =& \frac{1}{8} \left(N_3 \alpha_3^2 + N_1 \alpha_1^2\right) \alpha_2^2 \pi_2^2                               &=& \ \ \frac{1}{8}\ C_{2,31}\ C_{2,2} \\
		H_3 =& \frac{1}{8} \left(N_1 \alpha_1^2 + N_2 \alpha_2^2\right) \alpha_3^2 \pi_3^2                               &=& \ \ \frac{1}{8}\ C_{3,12}\ C_{3,3} \\
		H_4 =& \frac{1}{8} \left(N_2 \alpha_2^4 + \frac{2}{m_1} \alpha_2^2 \alpha_3^2 + N_3 \alpha_3^4\right) \pi_1^2    &=& \ \ \frac{1}{8}\ C_{4} \\
		H_5 =& \frac{1}{8} \left(N_3 \alpha_3^4 + \frac{2}{m_2} \alpha_3^2 \alpha_1^2 + N_1 \alpha_1^4\right) \pi_2^2    &=& \ \ \frac{1}{8}\ C_{5} \\
		H_6 =& \frac{1}{8} \left(N_1 \alpha_1^4 + \frac{2}{m_3} \alpha_1^2 \alpha_2^2 + N_2 \alpha_2^4\right) \pi_3^2    &=& \ \ \frac{1}{8}\ C_{6} \\
		H_7 =& -\frac{1}{4} \left(\frac{1}{m_3} \alpha_2 \pi_2 + \frac{1}{m_2} \alpha_3 \pi_3\right) \alpha_1^3 \pi_1    &=& -\frac{1}{4}\ C_{7,23}\ C_{7,1} \\
		H_8 =& -\frac{1}{4} \left(\frac{1}{m_1} \alpha_3 \pi_3 + \frac{1}{m_3} \alpha_1 \pi_1\right) \alpha_2^3 \pi_2    &=& -\frac{1}{4}\ C_{8,31}\ C_{8,2} \\
		H_9 =& -\frac{1}{4} \left(\frac{1}{m_2} \alpha_1 \pi_1 + \frac{1}{m_1} \alpha_2 \pi_2\right) \alpha_3^3 \pi_3    &=& -\frac{1}{4}\ C_{9,12}\ C_{9,3} ,
	\end{array}
\end{equation}
where each subindexed function $C_i$ is a constant of motion in its associated Hamiltonian.

There are clearly four groups in equation \eqref{eqn:splitting}, which we shall enumerate 0: \{0\}, 1: \{1,2,3\}, 2: \{4,5,6\} and 3: \{7,8,9\}. $H_0$ depends on coordinates only, so can be integrated by \eqref{eqn:porqonly}.
Group 1 can be integrated by \eqref{eqn:Hb},
group 2 can be integrated by \eqref{eqn:Ha}, 
and finally group 3 can be integrated by \eqref{eqn:Hc} where $G$ is a sum of $H_{mm}$ Hamiltonians.

\subsection{Higher order methods}
An important ingredient in constructing higher order reversible methods is the adjoint $(\phi^t)^*$  of a method
$\phi^t$ which is defined to be $(\phi^{-t})^{-1}$. 
If $\phi^t = \psi_1^t \circ \psi_2^t \circ \dots \circ \psi_n^t$ and each $\psi_i^t$ is self-adjoint, then 
the adjoint is obtained by reversing the order of composition. 
This follows from the definition of the adjoint: 
\begin{align*}
\label{eqn:adjoint}
(\phi^t)^* &= (\phi^{-t})^{-1} \\
                      &= (\psi_1^{-t}\circ \psi_2^{-t} \circ \dots \circ \psi_n^{-t})^{-1} \\
					  &= (\psi_n^{-t})^{-1} \circ (\psi_{n-1}^{-t})^{-1} \circ \dots \circ (\psi_1^{-t})^{-1} \\
					  &= \psi_n^t\circ \psi_{n-1}^t \circ \dots \circ \psi_1^t \,.
\end{align*}
In our case the self-adjointness of the individual steps $\psi_i^t$ follows from the fact that they are 
exact solution of Hamilton's equations. 

\label{subsec:higherorder}
Channell \& Neri 
\cite{ChannellNeri1996} offer a basic derivation of a reversible, symplectic map that is accurate to second order in the time step. When the splitting is of the form $H = T(p) + U(q)$ this leads to the symplectic leapfrog integrator, by composing symplectic Euler with its adjoint. 

This construction also applies to the more complicated case with a first order integrator composed of 10 self-adjoint maps as in our case. Given $\phi^t = \psi_1^t\circ \psi_2^t \circ \dots \circ \psi_n^t$ as above a reversible 
second order method is found as
\begin{align*}
	\phi_2^t &= \phi_1^{\frac{t}{2}} \circ ( \phi_1^{\frac{t}{2}} )^* \\ &= \psi_1^{\frac{t}{2}} \circ \dots \circ \psi_{n-1}^{\frac{t}{2}} \circ \psi_n^t\circ \psi_{n-1}^{\frac{t}{2}} \circ \dots \circ \psi_1^{\frac{t}{2}} .
\end{align*}

\cite{Yoshida1990} gives a general method by which one may obtain integrators of arbitrary even order, if only one has, to start with, a reversible even-order integrator $\phi_2^t$ such as the symplectic leapfrog---or, more generally, symplectic midpoint. One can compose $\phi_2^t$ to obtain a fourth order integrator $\phi_4^t$, and compose this to obtain $\phi_6^t$ and so on. In general, given $\phi_{2n}^t$,
\begin{equation}
	\label{eqn:higherorder}
	\phi_{2n+2}^t = \phi_{2n}^{z_1 t}\ \phi_{2n}^{z_0 t}\ \phi_{2n}^{z_1 t},
\end{equation}
where we define $z_0 = -\dfrac{2^{1/(2n+1)}}{2-2^{1/(2(2n+1))}}, z_1 = \dfrac{1}{2-2^{1/(2n+1)}}$ to adjust the step size of the lower order method.

{
This method is easy to construct and implement, but quickly becomes unwieldy. When $n = 2$ (order $4$), there are three evaluations of the second order method, but at orders $6$ and $8$ there are, respectively, nine and twenty-seven. As noted by
\cite{Yoshida1990}, there are better methods, and he gives coefficients for a sixth order method and several sets of coefficients for eighth order methods. The construction of higher order methods is discussed extensively in \cite{HaLuWa02} and \cite{mclachlan2002}. We will assess in section \ref{sec:numerics} which methods give good results for our problem comparing the methods whose coefficients are given in \cite{HaLuWa02} and those constructed by \cite{Yoshida1990}.

\subsection{Regularisation and variable time stepping}
\label{subsec:timestep}

There is a well known restriction on symplectic integration that such integrators must use a constant step size, or the benefits of these methods for large integration times are lost due to the introduction of new secular error terms. Various authors have discussed methods of achieving adaptive step size in symplectic integration that avoids this problem; for example, \cite{mikkola1997,pretoTremaine1999,blanesBudd2005,blanesIserles2012}.

In particular, \cite{blanesIserles2012} explore the use of Sundman and Poincar\'e transformations and give a good overview of the problem. In general the Sundman transformation is non-symplectic, though with care the transformation can be made to respect geometric structure. In the their framework, the time scaling $\deriv{t}{\tau} = a_1 a_2 a_3$ 
is called the \emph{monitor function}. 
Our situation is special because the regularisation transformation consists of two intimately related steps.
First there is the canonical extension of the transformation of coordinates from distances $a_j$ to their ``roots'' $\alpha_j$ (space regularisation),
and second there is the time scaling (time regularisation). The time scaling up to a constant factor is achieved 
using the square of the Jacobian determinant of the transformation of the coordinates.
Only the combination of the two achieves global regularisation. Treating the time transformation separately 
as a monitor function would mean to integrate singular equations, since the original equations are singular at collision,
and they are still singular after the spatial regularisation alone.
Slight modifications of the time scaling are possible, see the remark at the end of the next section.

In order to achieve variable time stepping a monitor function could be used in the way described by 
\cite{blanesIserles2012} by integrating another equation on top of the regularisation (in space and time)
we have done. This may be particularly useful when integrating orbits with large distances between the bodies.
} 

\subsection{Finite time blowup}
\label{subsec:blowup}
It must be noted that  the solution of the Hamiltonian $H =  q^m p^n$ given in \eqref{eqn:polysol1} can (for $n\ne m$) reach infinity in finite time. This occurs when
\begin{equation}
	\label{eqn:singularity}
	1 + (n-m) \ q_0^{m-1}\ p_0^{n-1} t = 0. \
\end{equation}

This obviously makes step sizes comparable to this threshold risky when this form of solution is used in the integrator. This singularity could be reached if the denominator is negative and large during forward timesteps, or if the denominator is positive and large for ``backward'' timesteps (as during the middle stage of Yoshida's trick).
This possibility arises in equation \eqref{eqn:K} in the group 4 of the splitting \eqref{eqn:splitting}, which have terms of the form $\alpha_j^3 \pi_j^1$.
It may appear that this finite time blowup is an artefact of the integrator. 
However, after the time scaling the Hamiltonian $K$ does have finite time blow up when particles escape to infinity.
In this light it seems less unexpected that a stage of the corresponding symplectic integrator shows the same behaviour.

\cite{Blanes2002} provides a means by which to avoid such singularities, by way of rewriting the polynomial in terms of sums of binomials in the coordinates and momenta and finding coefficients such that the two expressions are equal. 
We did apply this to the Hamiltonian $H_{31}$ in our problem, and found a way to replace this with a Cremona map.
However, it turned out that the overall error of the method was worse than without this modification. 
Our method is more expensive, since it needs to compute rational powers, but this additional cost 
is worth it.

{ 
A way to possibly avoid finite time blowup when the configuration of the system becomes large would be to consider a rational---rather than polynomial---time scaling function as in \cite{MoMo12}. One could consider, for example, $\deriv{t}{\tau} = \frac{a_1 a_2 a_3}{\alpha^{2\gamma}}$ (recalling $\alpha^2 = \sum \alpha_j^2$), which tends to $0$ for $\gamma = 3$ or to ${a_j}/{4}$ for $\gamma = 2$ as $\alpha_j \rightarrow \infty$. For negative energy, the only possible escape to infinity is of one single mass and a hard binary; in regularised coordinates this is exactly one coordinate tending to infinity while the other two remain bounded. 
Such a time scaling would inevitably require that the Hamiltonian be split differently, possibly with more stages and complexity. In principle the methods described in this paper apply as long as exact solutions can be found for the partial Hamiltonians.
Unfortunately we have not been able to solve all of the resulting rational Hamiltonians.
} 

\subsection{Other polynomial globally regularised Hamiltonians}

Our main concern in this paper is the zero-angular momentum reduced and regularised planar 3-body problem,
which has 3 degrees of freedom. Other well known globally regularised polynomial Hamiltonians are 
due to 
\cite{Waldvogel1972} for the planar 3-body problem and to 
\cite{Heggie74} for 
the spatial 3-body problem. These Hamiltonians are not fully symmetry reduced and have 4 degrees of freedom 
(planar arbitrary angular momentum) and 12 degrees of freedom (spatial arbitrary angular momentum).

Heggie's simultaneously regularised Hamiltonian for the spatial 3-body problem
\cite{Heggie74} in canonical variables $Q_{ji}$ and conjugate $P_{ji}$, $j = 1,2,3$, $i = 1,\dots,4$ has the form 
\begin{align*}
H &= H_0 + H_4 + H_5 + H_6 + H_{-1} \\
H_0 &=  - h a_1 a_2 a_3 - \sum m_j m_k a_j a_k, \\
H_{3 + l} &= \frac18 \frac{a_j a_k}{\mu_{jk}}  |p_l|^2, \quad l = 1, 2, 3\\
H_{-1} &= \frac{1}{4} \sum \frac{a_j}{m_j} ( A_k P_k ) \cdot  (A_l P_l) 
\end{align*}
where $\mu_{jk} = m_j m_k / (m_j + m_k)$, and $a_j = \sum_{i  = 1}^4 Q_{ji}^2$, $|p_j|^2 =  \sum_{i  = 1}^4 P_{ji}^2$.
In addition $P_l = (P_{l1}, P_{l2}, P_{l3}, P_{l4})^T$ and $A_l$ is the KS-matrix \cite{kustaanheimo65} of the form 
\[
  A_l = \begin{pmatrix}
     Q_{l1} & -Q_{l2} & -Q_{l3} & Q_{l4} \\
     Q_{l2} & Q_{l1} & -Q_{l4} & -Q_{l3} \\
      Q_{l3} & Q_{l4} & Q_{l1} & Q_{l2} \\
  \end{pmatrix}
\]
The terms $H_0$ and $H_{4,5,6}$ are analogous to the previous ones.
The terms in $H_{-1}$ can be split into 9 terms of the form $a_j f_k g_l$ where 
the functions $f_k$ and $g_l$ only depend on the degrees of freedom $k$ and $l$, respectively.
These terms are somewhat similar to the Hamiltonians $H_{1,2,3}$ in Waldvogel's case.
Thus the Hamiltonian can be split into 13 polynomials of degree up to 6 each of which is integrable.

When setting $Q_{ji}$ and $P_{ji}$ with $i=3,4$ equal to zero Heggie's Hamiltonian describes a planar problem.
However, this still has 6 degrees of freedom. We can reduce the number of degrees of freedom to 4 
by instead using Waldvogel's Hamiltonian \cite{Waldvogel1972,orbitsin3bp}. 
This Hamiltonian is a polynomial of degree 12 and can be split into 15 terms
in a way similar to the two cases discussed above.

\section{Numerical examples}

\label{sec:numerics}
In this section we will show some numerical results achieved using our integrator in a selection of orbits ranging from far from collision to close encounters to a collision orbit.

\begin{figure}[t]
	\centering
	\includegraphics[width=1.0\textwidth]{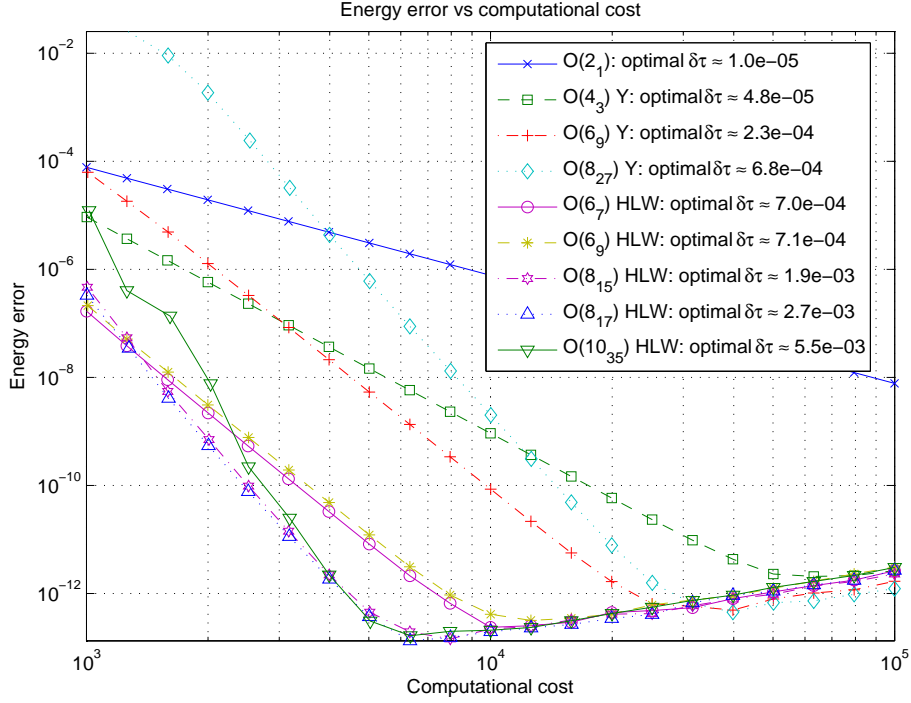}
	\caption{{Averaged error, integrating a fixed time interval for varying integration costs. Time step size at the minimal error is listed for each order, as well as our source for the method.}}
	\label{fig:averror}
\end{figure}

Figure \ref{fig:averror} shows 
{ the energy error for various integration methods in a ``work-precision'' diagram.
The error is averaged over several different initial conditions integrated over a fixed time interval.
The error is displayed as a function of the computational cost. The methods compared are 
the base method of order $2_1$, the integrators of \cite{Yoshida1990} ($4_3$, $6_9$, and $8_{27}$)
and other higher order symmetric compositions of symmetric methods of various authors, 
whose coefficients are given in \cite{HaLuWa02}, 
section~V.3.2, also see the references therein. The subscript with each method's order indicates the number of second order substeps in the evaluation of a single time step, indicating the cost of each method, where the second order method is given the base cost of 1.
A close look at the graph reveals that integrator $8_{17}$ achieves the lowest error with a step size of about $0.0027$, though it is a close call between any of the  three best methods $8_{15}$, $8_{17}$ and $10_{35}$. Reducing the step size further creates larger round-off errors.
All of the following examples are calculated with the $8_{17}$ integrator and step size $0.0027$, unless otherwise mentioned.
} 

\begin{figure}[t]
	\centering
	\begin{subfigure}[b]{\wi\textwidth}
		\centering
		\includegraphics[width=\textwidth]{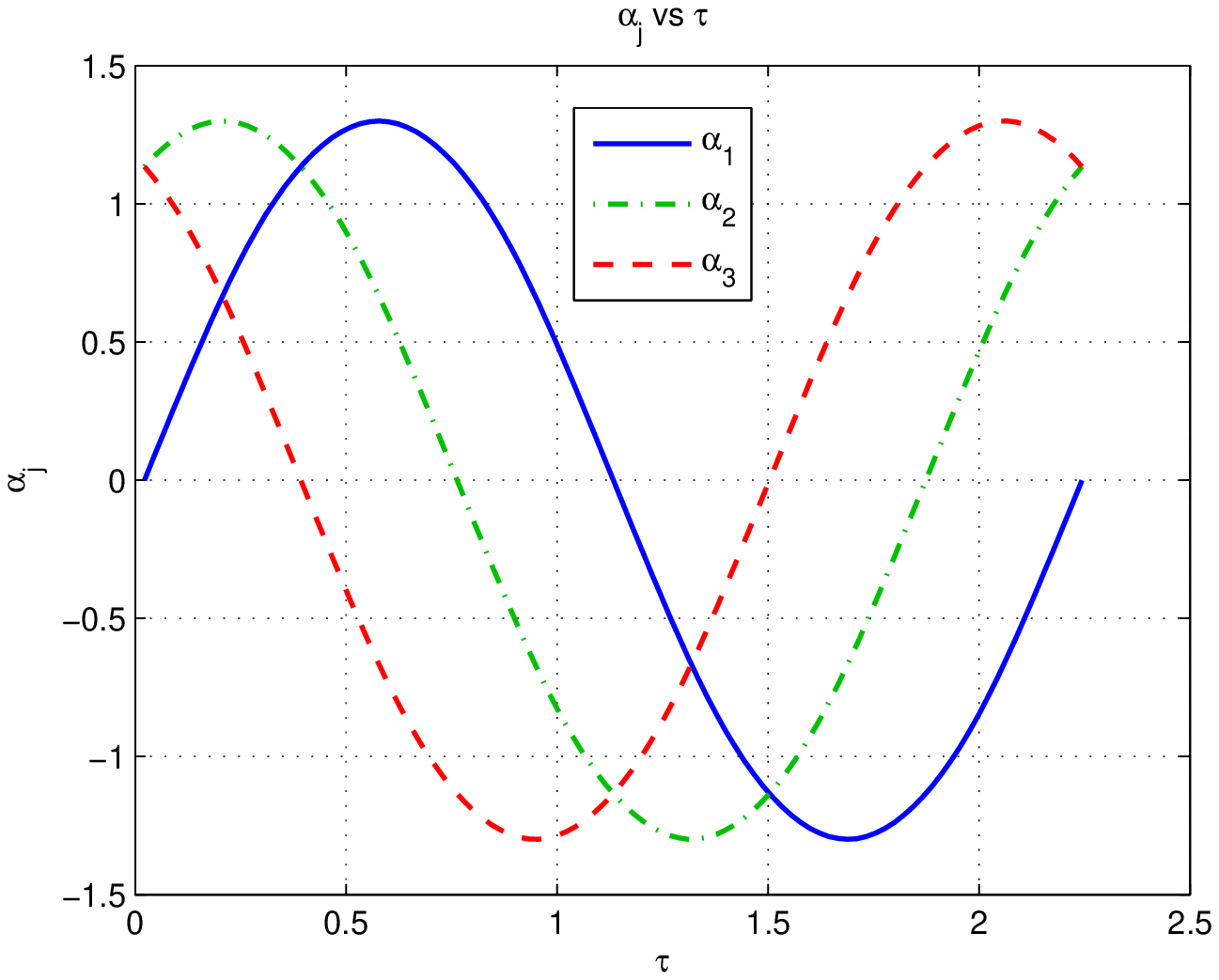}
		\caption{Regularised coordinates.}
		\label{fig:fig8alpha}
	\end{subfigure}
	~
	\begin{subfigure}[b]{\wi\textwidth}
		\centering
		\includegraphics[width=\textwidth]{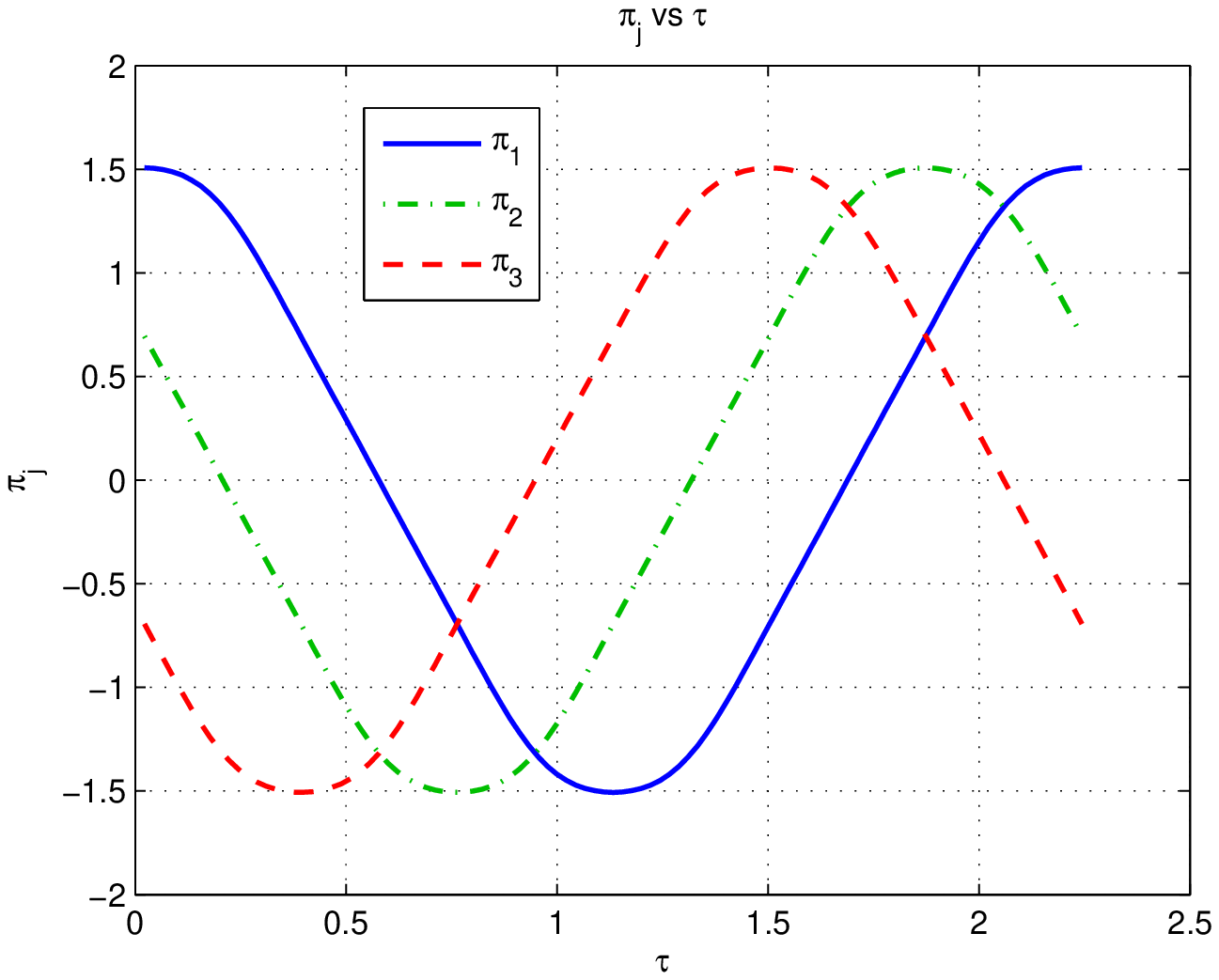}
		\caption{Regularised momenta.}
		\label{fig:fig8pi}
	\end{subfigure}
	~
	\begin{subfigure}[b]{\wi\textwidth}
		\centering
		\includegraphics[width=\textwidth]{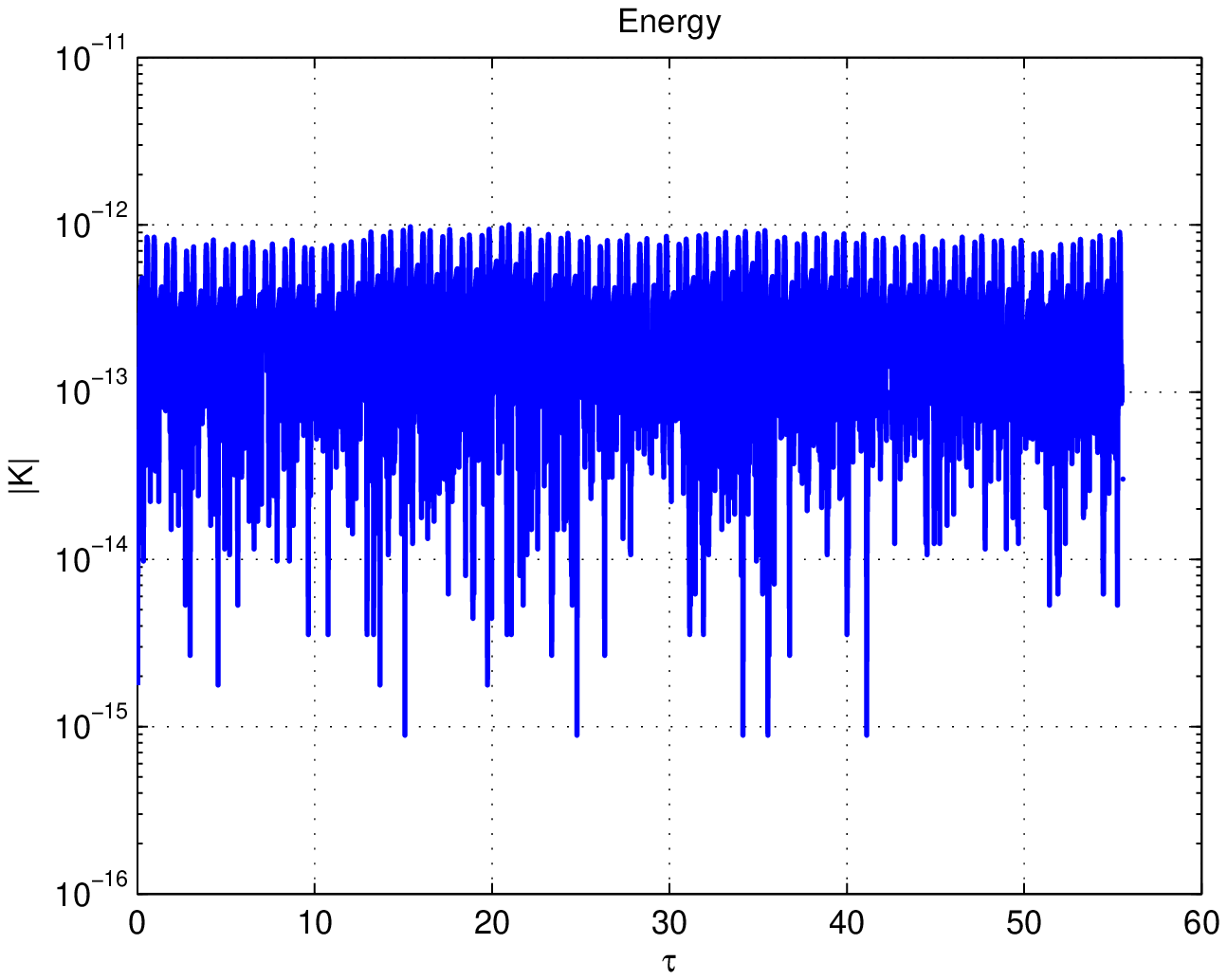}
		\caption{Energy error for 25 periods with 200 time steps per period.}
		\label{fig:fig8K}
	\end{subfigure}
	~
	\begin{subfigure}[b]{\wi\textwidth}
		\centering
		\includegraphics[width=\textwidth]{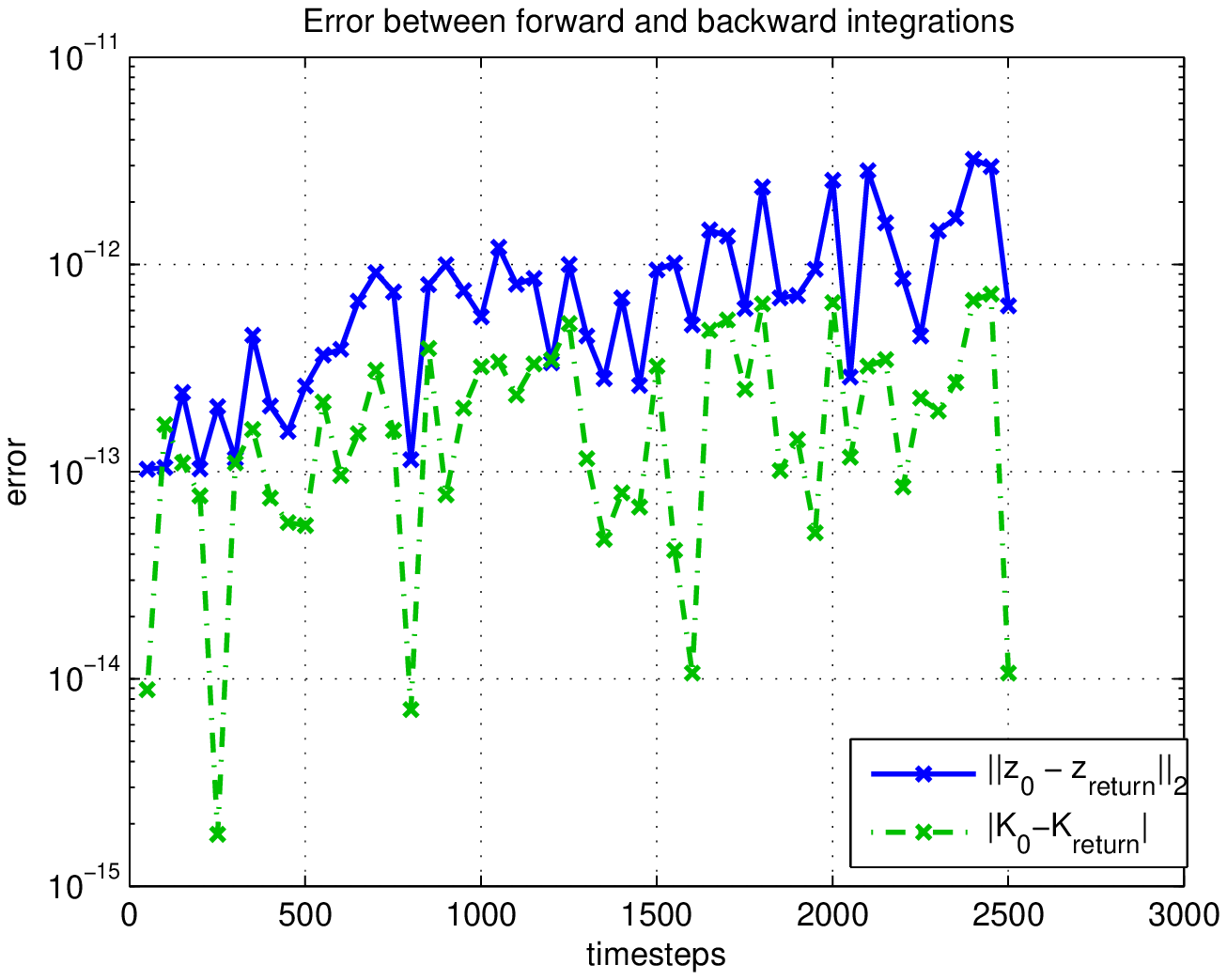}
		\caption{Two-way integration error over 25 periods.}
		\label{fig:fig8error}
	\end{subfigure}
	~
	\caption{The figure--8 choreography with scaled period $2{.}221813718$.}
	\label{fig:fig8details}
\end{figure}

\begin{figure}
	\centering
	\includegraphics[width=.95\textwidth]{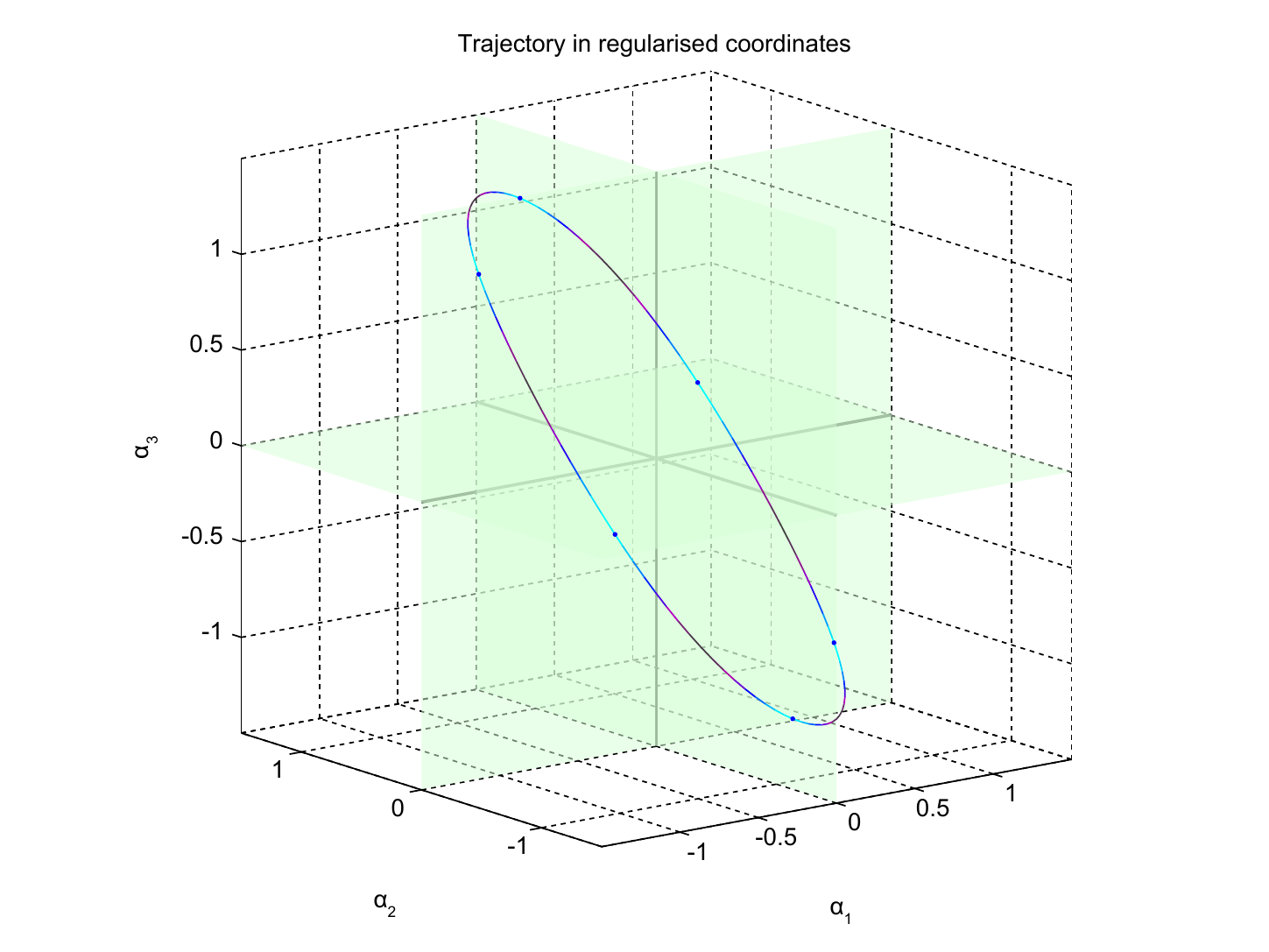}
	\caption{Trajectory of figure---8 choreography in $\alpha$-space, lying nearly in a plane. Colour gradient represents the moment of intertia (lighter is higher).}
	\label{fig:fig8-3d}
\end{figure}

Consider the figure--8 choreography, discovered by 
\cite{fig8paper0Moore}, proved to exist by 
\cite{fig8paper1ChencinerMontgomery} and explored by 
\cite{fig8paper2Simo,fig8paper3Simo}. We choose initial conditions
\begin{align*}
	\bb{\alpha}_0 &= (  0, 1.134522804969261, 1.134522804969261 )^T \\
	\bb{\pi}_0    &= ( 1.506773685132772, 0.694233777317562, -0.694233777317562 )^T
\end{align*}
in regularised coordinates, with $h = -1$ and equal unit masses. 
In scaled time, the figure-8 has a period of $2{.}221813718$; in physical time its period is $9.2371333$. The trajectory in regularised coordinates is shown in figures \ref{fig:fig8alpha} and \subref{fig:fig8pi} and the energy error over 25 orbits 
{with large time steps given by the period divided by 200} is shown in figure \ref{fig:fig8K}. Figure \ref{fig:fig8-3d} shows the trajectory of this orbit in the 3-dimensional space $(\alpha_j)$. Note that each crossing of a plane $\alpha_j = 0$ corresponds to a syzygy with $m_j$ in the middle of the configuration.

\begin{figure}[t]
	\centering
	\begin{subfigure}[b]{\wi\textwidth}
		\centering
		\includegraphics[width=\textwidth]{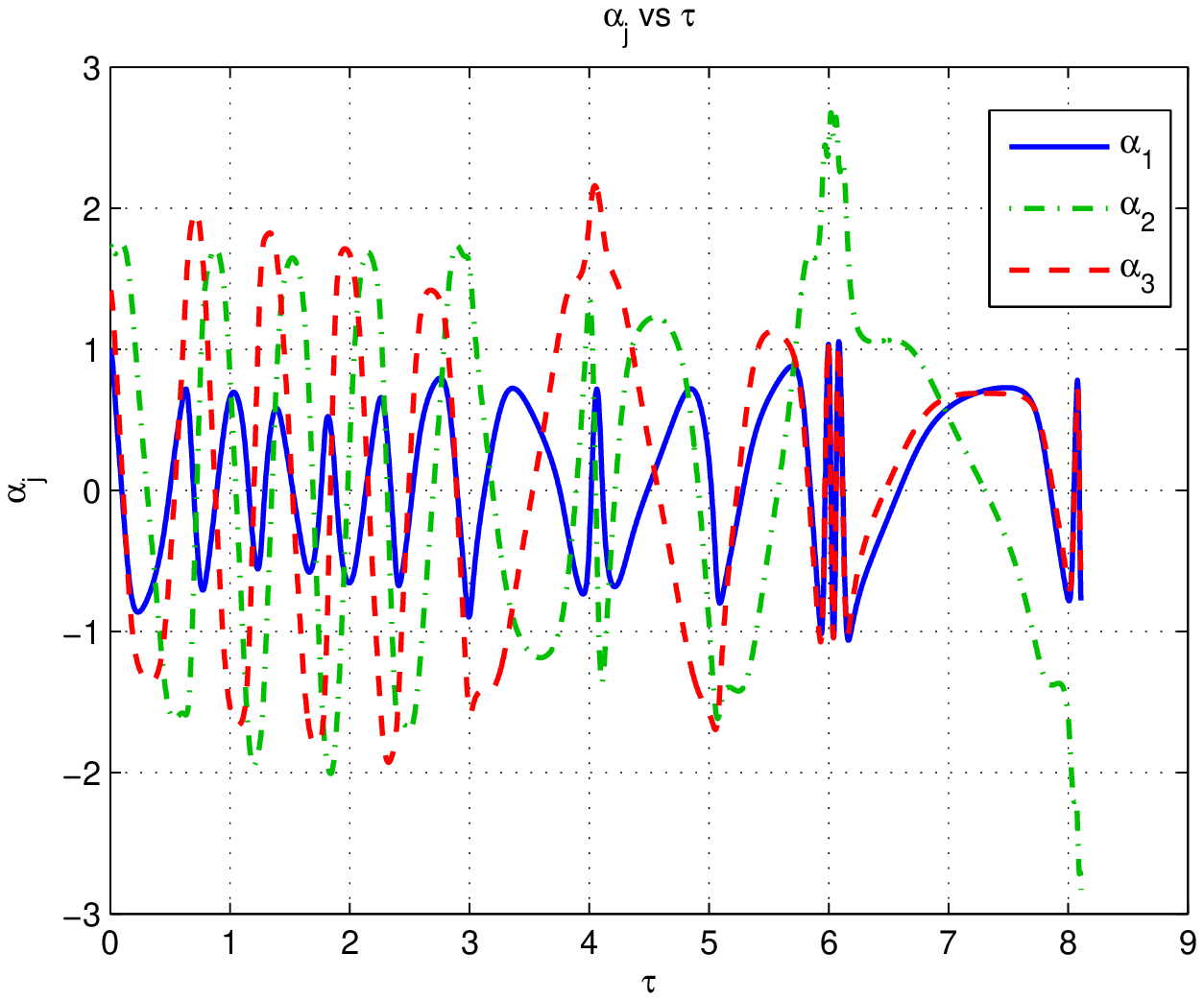}
		\caption{Regularised coordinates.}
		\label{fig:pythalpha}
	\end{subfigure}
	~
	\begin{subfigure}[b]{\wi\textwidth}
		\centering
		\includegraphics[width=\textwidth]{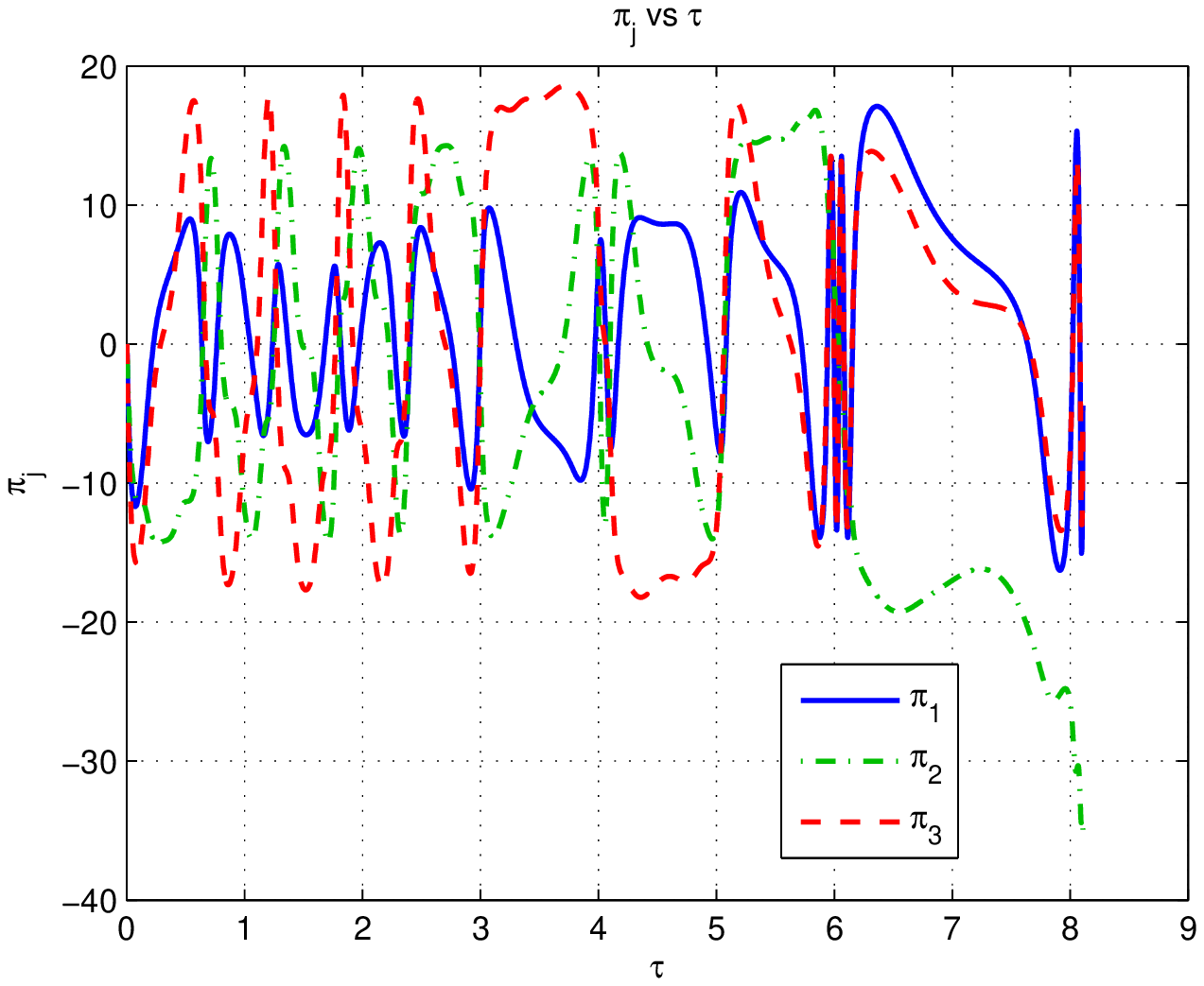}
		\caption{Regularised momenta.}
		\label{fig:pythpi}
	\end{subfigure}
	~
	\begin{subfigure}[b]{\wi\textwidth}
		\centering
		\includegraphics[width=\textwidth]{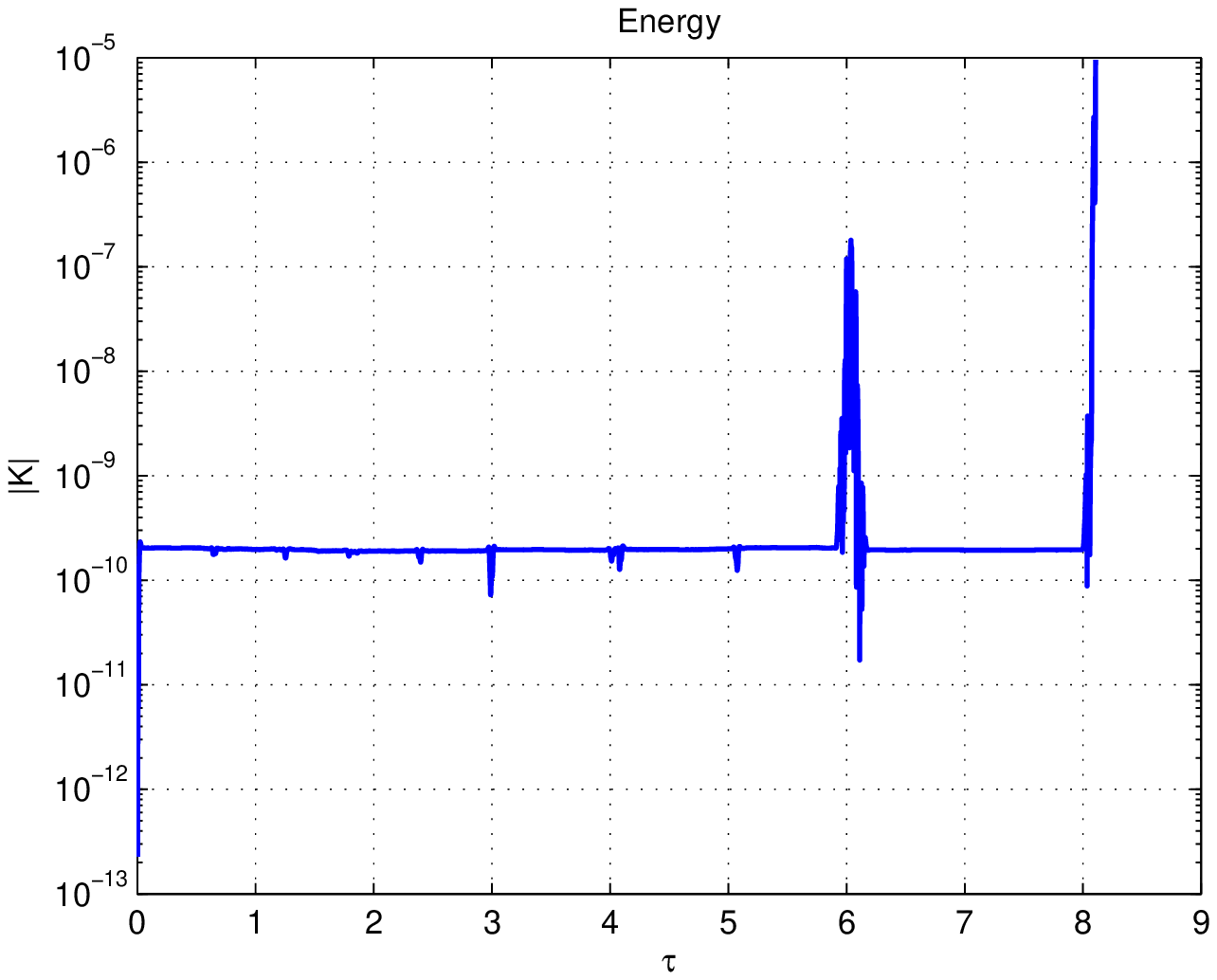}
		\caption{Energy error.}
		\label{fig:pythK}
	\end{subfigure}
	~
	\begin{subfigure}[b]{\wi\textwidth}
		\centering
		\includegraphics[width=\textwidth]{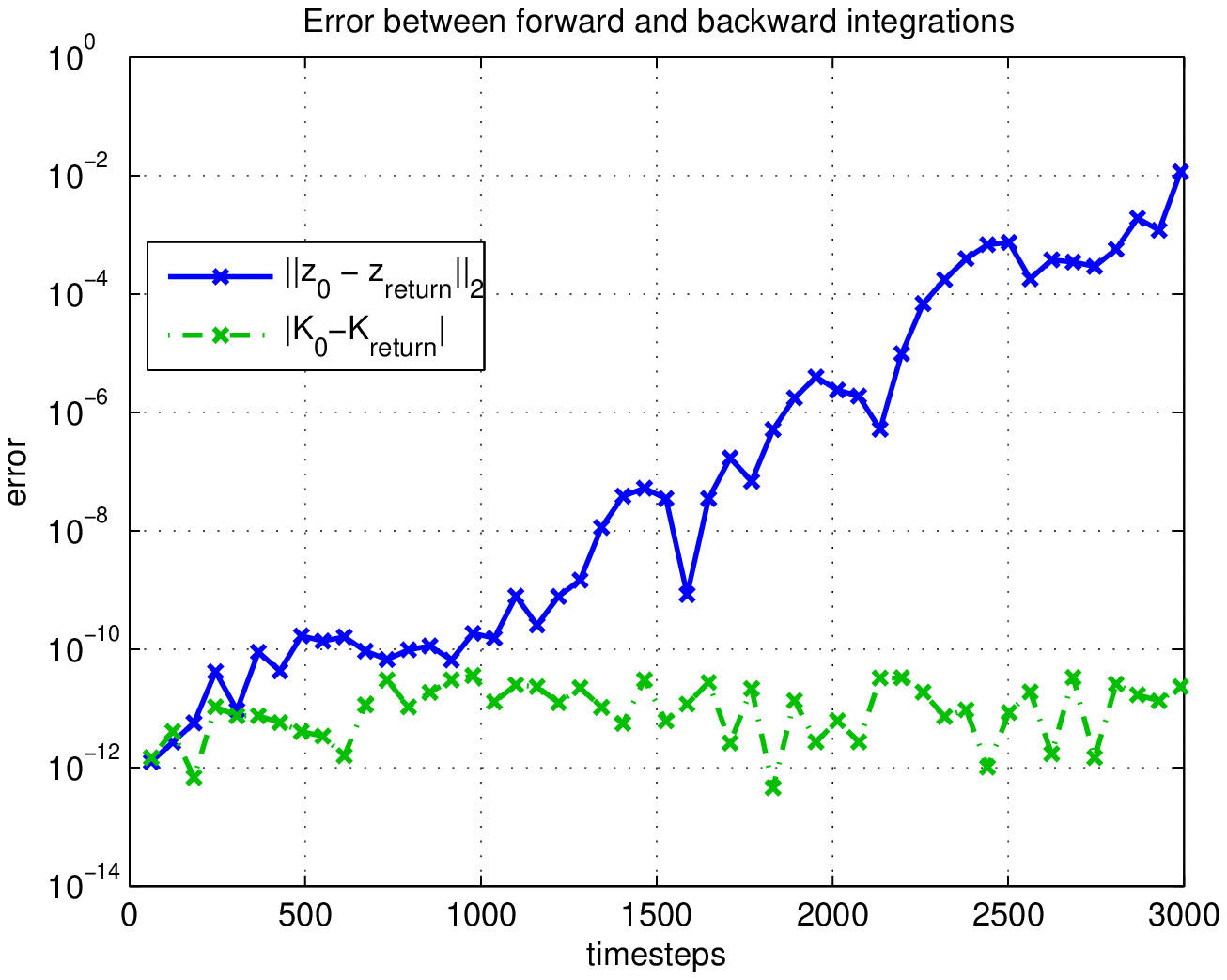}
		\caption{Two-way integration error.}
		\label{fig:pytherror}
	\end{subfigure}
	~
	\caption{The Pythagorean orbit integrated up to $\tau=8{.}105$ in scaled time.}
	\label{fig:pythdetails}
\end{figure}

Next we look at the Pythagorean orbit \cite{pythagOrbit1967} for $m_1=3$, $m_2=4$, $m_3 = 5$, 
with initial conditions as given in \cite{orbitsin3bp}, which, in regularised coordinates, are
\begin{align*}
	\bb{\alpha}_0 &= ( 1 , \sqrt{3} , \sqrt{2})^T \\
	\bb{\pi}_0    &=( 0,  0, 0)^T.
\end{align*}
This orbit has a close encounter between masses 1 and 3 at around $t = 15.8$ in physical time (about $\tau = 1.52$ in scaled time). 
Waldvogel's analysis regularises the system, albeit slightly differently, and his integration is not symplectic. The final motions of this orbit compare well with other studies; plotting the orbit in physical space produces results indistinguishable from \cite{pythagOrbit1967}, \cite{orbitsin3bp}.

The regularisation of the 3---body problem allows our integrator to cope well when the distances between any two masses are small. The result of the time scaling is that the regularised system has a finite time blowup for any escape orbit. If one continues to integrate the Pythagorean orbit past $\tau = 8.105$, the error in the energy grows exponentially and the results become inaccurate.

\begin{figure}[t]
	\centering
	\begin{subfigure}[b]{\wi\textwidth}
		\centering
		\includegraphics[width=\textwidth]{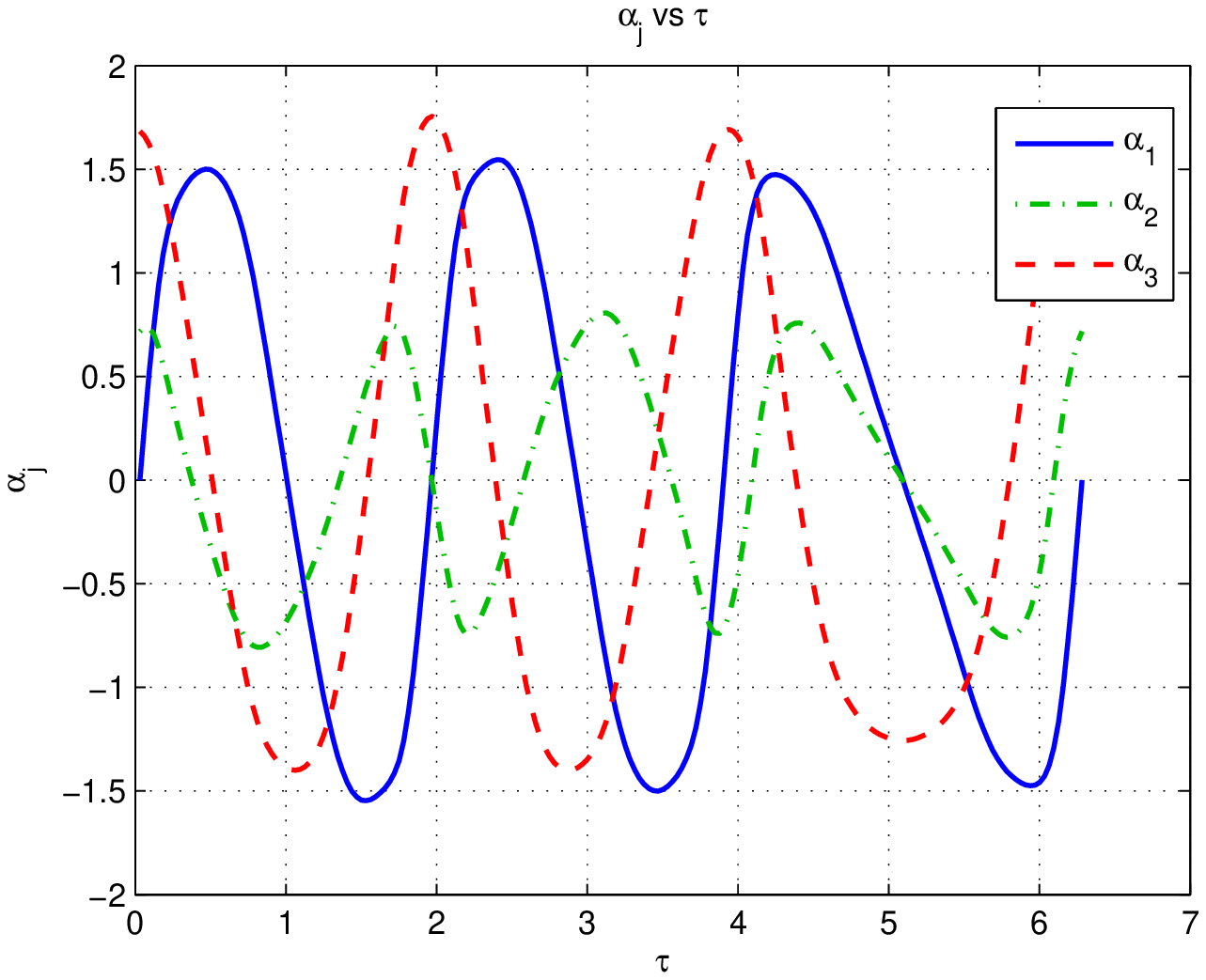}
		\caption{Regularised coordinates.}
		\label{fig:collalpha}
	\end{subfigure}
	~
	\begin{subfigure}[b]{\wi\textwidth}
		\centering
		\includegraphics[width=\textwidth]{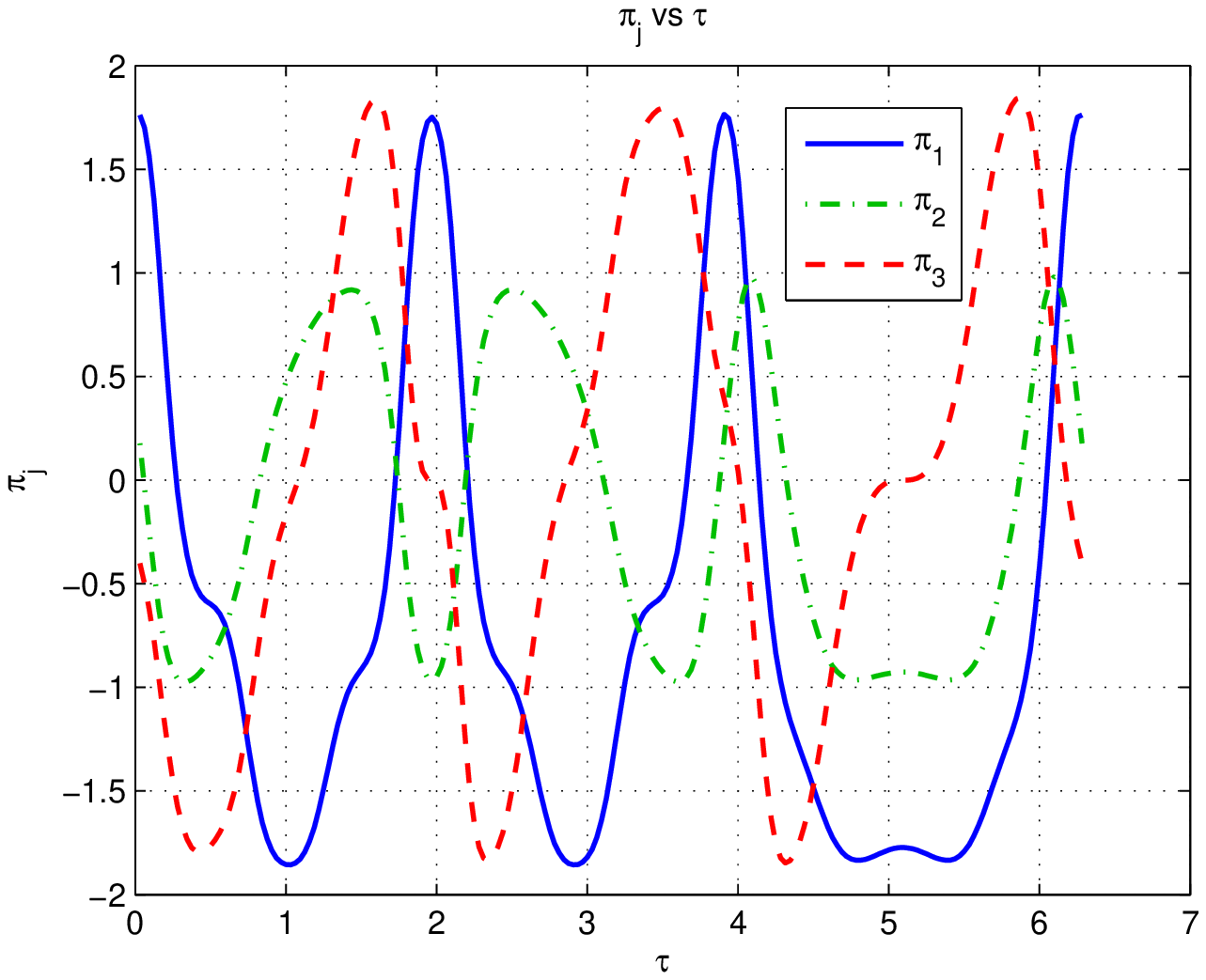}
		\caption{Regularised momenta.}
		\label{fig:collpi}
	\end{subfigure}
	~
	\begin{subfigure}[b]{\wi\textwidth}
		\centering
		\includegraphics[width=\textwidth]{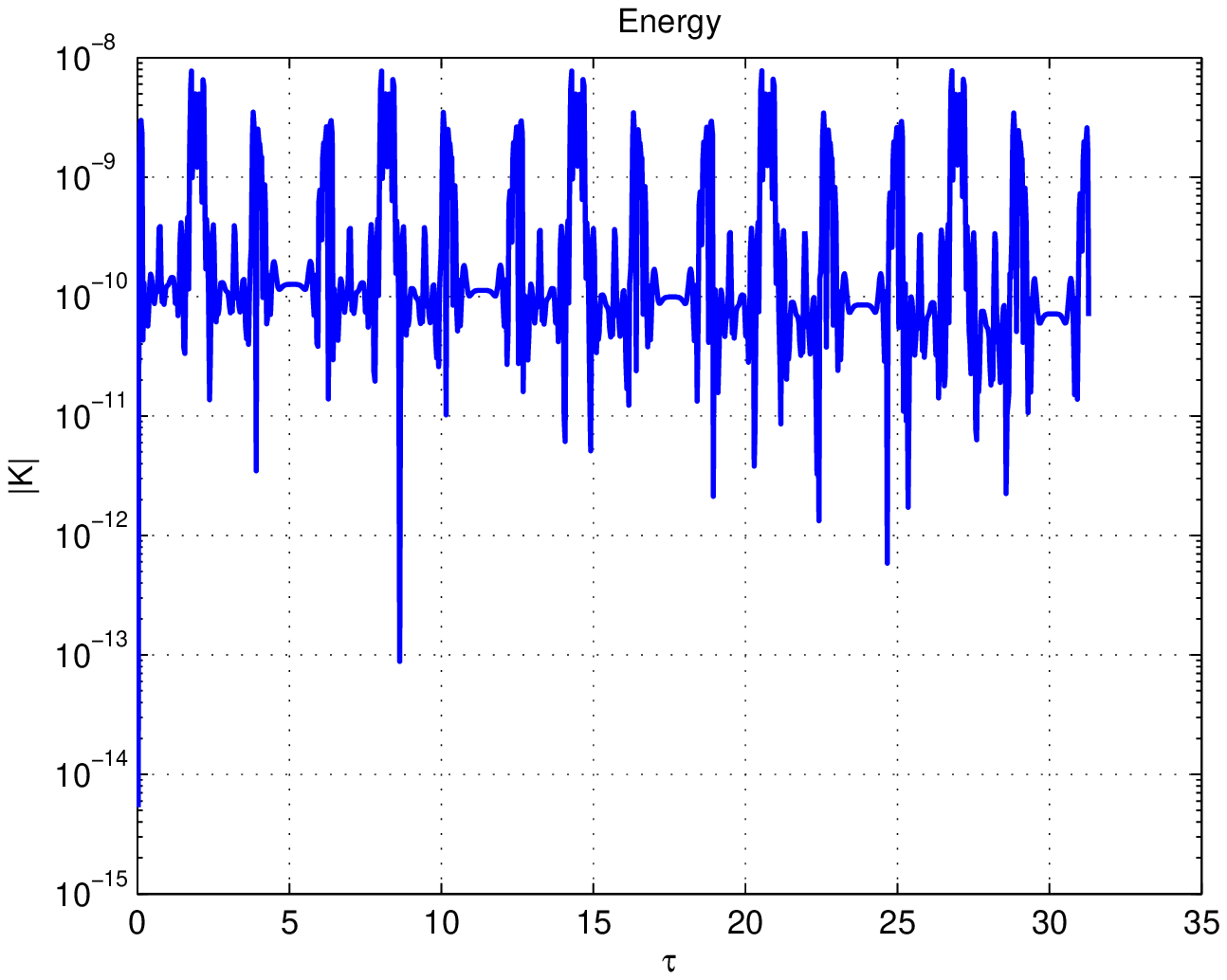}
		\caption{Energy error for 5 periods.}
		\label{fig:collK}
	\end{subfigure}
	~
	\begin{subfigure}[b]{\wi\textwidth}
		\centering
		\includegraphics[width=\textwidth]{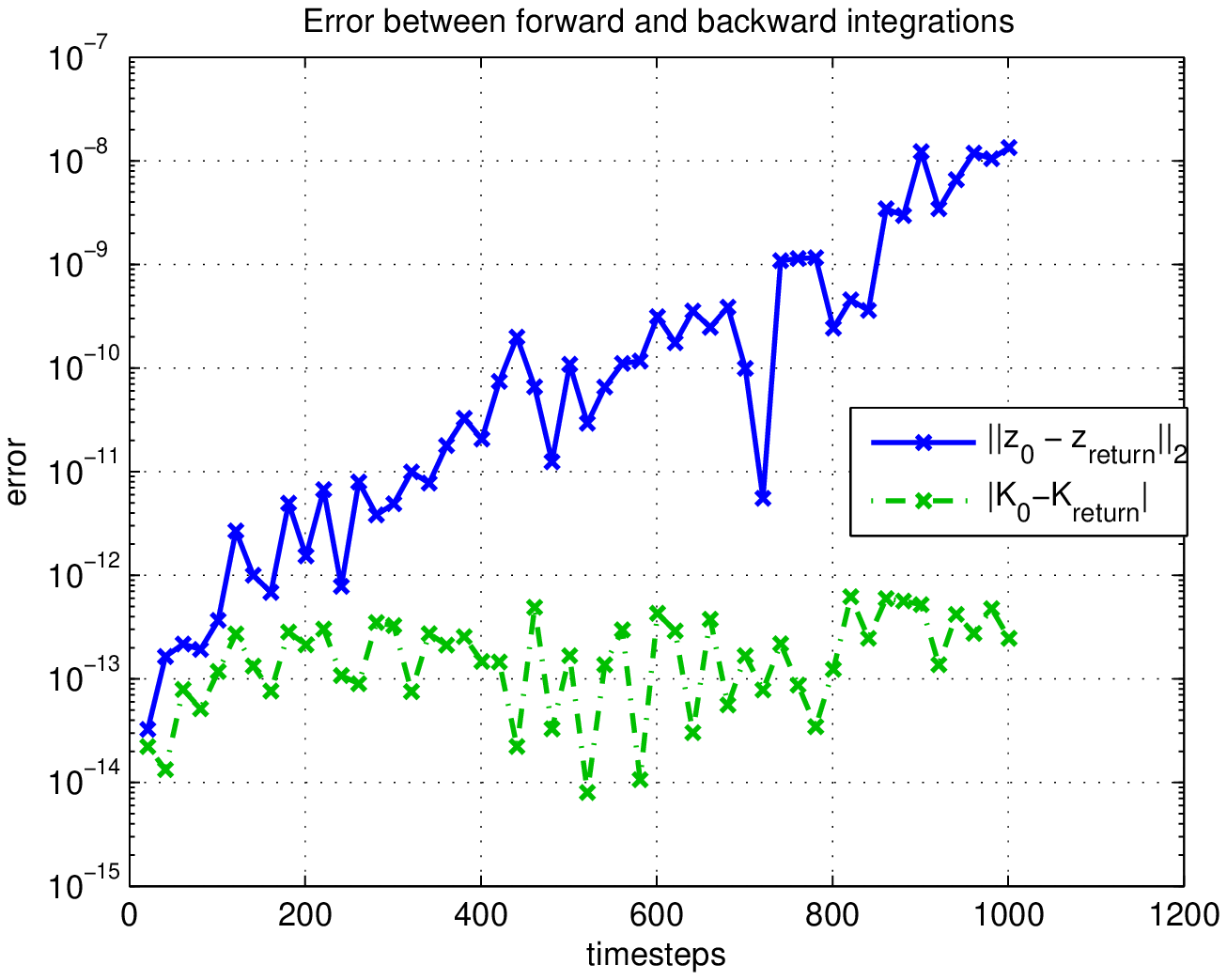}
		\caption{Two-way integration error for 5 periods.}
		\label{fig:collerror}
	\end{subfigure}
	~
	\caption{A periodic collision orbit with scaled period  $6{.}2520511$ with 200 time steps per period.}
	\label{fig:colldetails}
\end{figure}

\begin{figure}
	\centering
	\includegraphics[width=.95\textwidth]{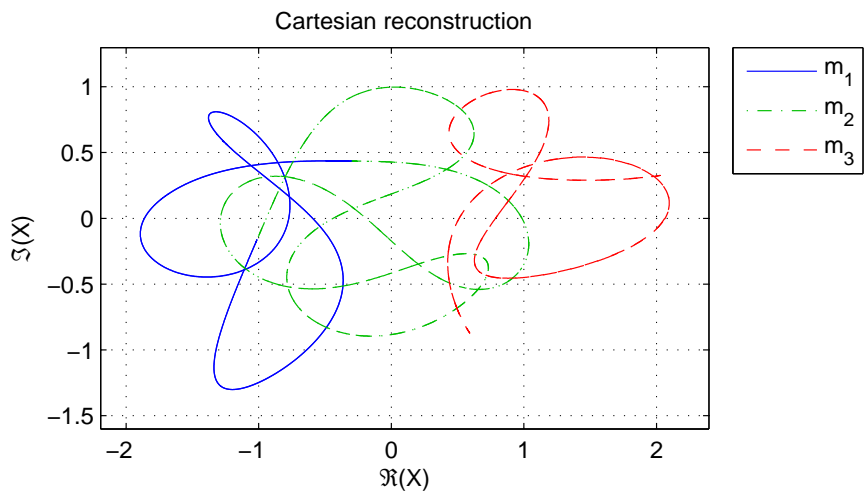}
	\caption{Reconstruction of Cartesian trajectories from regularised integration for a periodic collision orbit.}
	\label{fig:collcartesian}
\end{figure}

\begin{figure}
	\centering
	\includegraphics[width=.85\textwidth]{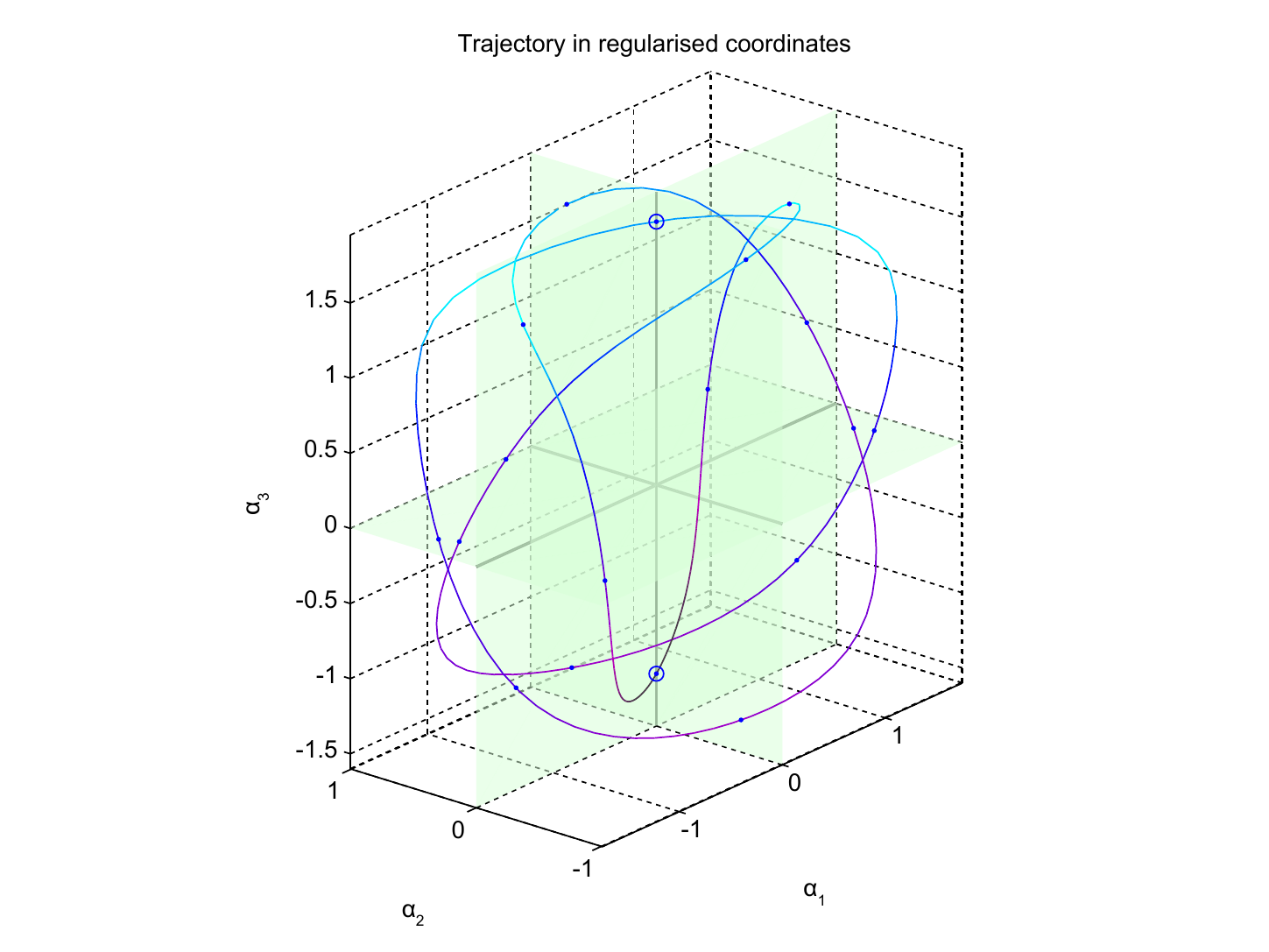}
	\caption{Trajectory of periodic collision orbit in $\alpha$-space. Colour gradient represents the moment of intertia of the configuration (lighter is higher). { The two big dots mark the (regularised) collisions.}}
	\label{fig:coll3d}
\end{figure}

Finally, we show results in a periodic collision orbit, discovered during a search for periodic orbits in the reduced space, with initial conditions
\begin{align*}
	\bb{\alpha}_0 &= ( 0 , 0.717162073833634 , 1.683647749751810 )^T\\
	\bb{\pi}_0    &=( 1.762174970761679, 0.177158588505747, -0.401743282150556 )^T,
\end{align*}
equal unit masses and $h = -1$. 
This orbit has {two} collisions between masses 1 and 2, as can be seen by $\alpha_1 = \alpha_2 = 0$ at $\tau = 1.9362$ and $\tau = 5.062$ in figure \ref{fig:collalpha}. This orbit {is periodic in full phase space and} is shown in figures \ref{fig:colldetails}, \ref{fig:collcartesian} and \ref{fig:coll3d}. Its scaled period is $6.2520511$, corresponding to a physical period of $29.6117209$. Note in figure \ref{fig:coll3d} that the collisions happen on the $\alpha_3$-axis, when $\alpha_1 = \alpha_2 = 0$.

Because Hamiltonian systems are time-reversible, it is desirable to have an integrator with the same property. The second order map $\phi_2^t$ is constructed as such, and Yoshida's formula for higher order integrators constructs them to be reversible as well. That means that $\phi_{2n}^t\ \phi_{2n}^{-t} = Id$ up to roundoff error. Figures \ref{fig:fig8error}, \ref{fig:pytherror} and \ref{fig:collerror} show how closely this integrator returns to its initial condition after a certain number of time steps in one direction, followed by the same number of iterations with a negative time step.

By this measure, the integrator has the most trouble with Pythagorean orbit, which clearly shows signs that it exists within a chaotic region of phase space by the exponential growth of error. However; energy is well preserved in this and the other cases.

\clearpage

\section{Conclusion}

We have constructed a symplectic integrator for the reduced and regularised planar 3-body problem 
at zero angular momentum. The method works well, but it is not very efficient, because each (first order) 
time step involves the computation of 10 individual maps. Our interests is the computation of relative 
periodic orbits including collision orbits, and for this task the method is appropriate. The detailed
results about relative periodic orbits and their geometric phase will be reported in a forthcoming paper.

\begin{appendices}
\section{Integrator stages}
\label{app:stages}
Subsection \ref{subsec:monomialintegration} described how to integrate a monomial Hamiltonian, and subsection \ref{subsec:splitting} described the splitting of equation \eqref{eqn:K} into a minimal number of solvable parts and those solutions. Here we use those solutions to build an explicit first order symplectic composition method for \eqref{eqn:K}.

Let the timestep be $\Dtau$, let $\mu_j = (m_k + m_l)$, let the values of the system before and after one timestep respectively be $\bb{z}_0 = (\alpha_{1,0},\dots,\pi_{3,0})^T$ and $\bb{z}_1 = (\alpha_{1,1},\dots,\pi_{3,1})^T$ and intermediate steps be $\bb{\xi}_i = (\alpha_{1,.i-1},\dots,\pi_{3,.i-1})^T$. Now
\begin{align*}
\label{eqn:stages}
\bb{\xi}_1 =& 
	\begin{pmatrix}
		\alpha_{1,0} \\
		\alpha_{2,0} \\
		\alpha_{3,0} \\
		\pi_{1,0} + 2 \alpha_{1,0} \left(\left(2\alpha_{1,0}^2 + a_{1,0}\right)\left(h a_{1,0} + M_1\right) + m_1 \alpha_{1,0} \mu_1 a_{1,0}\right) \Dtau \\
		\pi_{2,0} + 2 \alpha_{2,0} \left(\left(2\alpha_{2,0}^2 + a_{2,0}\right)\left(h a_{2,0} + M_2\right) + m_2 \alpha_{2,0} \mu_2 a_{2,0}\right) \Dtau \\
		\pi_{3,0} + 2 \alpha_{3,0} \left(\left(2\alpha_{3,0}^2 + a_{3,0}\right)\left(h a_{3,0} + M_3\right) + m_3 \alpha_{3,0} \mu_3 a_{3,0}\right) \Dtau
	\end{pmatrix}\\
\bb{\xi}_2 =&
\begin{pmatrix}
	\alpha_{1,.1}\ \exp \left( \frac{1}{4} \left(N_2 \alpha_{2,.1}^2 + N_3 \alpha_{3,.1}^2\right) \alpha_{1,.1}\ \pi_{1,.1}\ \Dtau \right) \\
	\alpha_{2,.1} \\
	\alpha_{3,.1} \\
	\pi_{1,.1}\ \exp \left( -\frac{1}{4} \left(N_2 \alpha_{2,.1}^2 + N_3 \alpha_{3,.1}^2\right) \alpha_{1,.1}\ \pi_{1,.1}\ \Dtau \right) \\
	\pi_{2,.1} - \frac{1}{4}\ N_2\ \alpha_{1,.1}^2\ \pi_{1,.1}^2\ \alpha_{2,.1} \Dtau \\
	\pi_{3,.1} - \frac{1}{4}\ N_3\ \alpha_{1,.1}^2\ \pi_{1,.1}^2\ \alpha_{3,.1} \Dtau
\end{pmatrix} \\
\bb{\xi}_3 =&
	\begin{pmatrix}
		\alpha_{1,.2} \\
		\alpha_{2,.2}\ \exp \left( \frac{1}{4} \left(N_3 \alpha_{3,.2}^2 + N_1 \alpha_{1,.2}^2\right) \alpha_{2,.2}\ \pi_{2,.2}\ \Dtau \right) \\
		\alpha_{3,.2} \\
		\pi_{1,.2} - \frac{1}{4}\ N_1\ \alpha_{2,.2}^2\ \pi_{2,.2}^2\ \alpha_{1,.2} \Dtau \\
		\pi_{2,.2}\ \exp \left( -\frac{1}{4} \left(N_3 \alpha_{3,.2}^2 + N_1 \alpha_{1,.2}^2\right) \alpha_{2,.2}\ \pi_{2,.2}\ \Dtau \right) \\
		\pi_{3,.2} - \frac{1}{4}\ N_3\ \alpha_{2,.2}^2\ \pi_{2,.2}^2\ \alpha_{3,.2} \Dtau
	\end{pmatrix} \\
\bb{\xi}_4 =&
	\begin{pmatrix}
		\alpha_{1,.3} \\
		\alpha_{2,.3} \\
		\alpha_{3,.3}\ \exp \left( \frac{1}{4} \left(N_1 \alpha_{1,.3}^2 + N_2 \alpha_{2,.3}^2\right) \alpha_{3,.3}\ \pi_{3,.3}\ \Dtau \right) \\
		\pi_{1,.3} - \frac{1}{4}\ N_1\ \alpha_{3,.3}^2\ \pi_{3,.3}^2\ \alpha_{1,.3} \Dtau \\
		\pi_{2,.3} - \frac{1}{4}\ N_2\ \alpha_{3,.3}^2\ \pi_{3,.3}^2\ \alpha_{2,.3} \Dtau \\
		\pi_{3,.3}\ \exp \left( -\frac{1}{4} \left(N_1 \alpha_{1,.3}^2 + N_2 \alpha_{2,.3}^2\right) \alpha_{3,.3}\ \pi_{3,.3}\ \Dtau \right)
	\end{pmatrix} \\
\bb{\xi}_5 =&
	\begin{pmatrix}
		\alpha_{1,.4} + \pi_{1,.4} \left(\frac{1}{4} \left(N_2 \alpha_{2,.4}^4 + N_3 \alpha_{3,.4}^4\right) + \frac{1}{2 m_1}\alpha_{2,.4}^2 \alpha_{3,.4}^2\right) \Dtau \\
		\alpha_{2,.4} \\
		\alpha_{3,.4} \\
		\pi_{1,.4} \\
		\pi_{2,.4} + \frac{1}{2} \pi_{1,.4}^2 \left(N_2 \alpha_{2,.4}^3 + \frac{1}{m_1} \alpha_{2,.4} \alpha_{3,.4}^2\right) \Dtau \\
		\pi_{3,.4} + \frac{1}{2} \pi_{1,.4}^2 \left(N_3 \alpha_{3,.4}^3 + \frac{1}{m_1} \alpha_{3,.4} \alpha_{2,.4}^2\right) \Dtau
	\end{pmatrix} \\
\bb{\xi}_6 =&
	\begin{pmatrix}
		\alpha_{1,.5} \\
		\alpha_{2,.5} + \pi_{2,.5} \left(\frac{1}{4} \left(N_3 \alpha_{3,.5}^4 + N_1 \alpha_{1,.5}^4\right) + \frac{1}{2 m_2}\alpha_{3,.5}^2 \alpha_{1,.5}^2\right) \Dtau \\
		\alpha_{3,.5} \\
		\pi_{1,.5} + \frac{1}{2} \pi_{2,.5}^2 \left(N_1 \alpha_{1,.5}^3 + \frac{1}{m_2} \alpha_{1,.5} \alpha_{3,.5}^2\right) \Dtau \\
		\pi_{2,.5} \\
		\pi_{3,.5} + \frac{1}{2} \pi_{2,.5}^2 \left(N_3 \alpha_{3,.5}^3 + \frac{1}{m_2} \alpha_{3,.5} \alpha_{1,.5}^2\right) \Dtau
	\end{pmatrix} \\
\bb{\xi}_7 =&
	\begin{pmatrix}
		\alpha_{1,.6} \\
		\alpha_{2,.6} \\
		\alpha_{3,.6} + \pi_{3,.6} \left(\frac{1}{4} \left(N_1 \alpha_{1,.6}^4 + N_2 \alpha_{2,.6}^4\right) + \frac{1}{2 m_3}\alpha_{1,.6}^2 \alpha_{2,.6}^2\right) \Dtau \\
		\pi_{1,.6} + \frac{1}{2} \pi_{3,.6}^2 \left(N_1 \alpha_{1,.6}^3 + \frac{1}{m_3} \alpha_{1,.6} \alpha_{2,.6}^2\right) \Dtau \\
		\pi_{2,.6} + \frac{1}{2} \pi_{3,.5}^2 \left(N_2 \alpha_{2,.5}^3 + \frac{1}{m_3} \alpha_{2,.5} \alpha_{1,.5}^2\right) \Dtau \\
		\pi_{3,.6}
	\end{pmatrix} \\
\bb{\xi}_8 =&
	\begin{pmatrix}
		\alpha_{1,.7} \left(1 + \frac{1}{2} \left(\frac{1}{m_3} \alpha_{2,.7} \pi_{2,.7} + \frac{1}{m_2} \alpha_{3,.7} \pi_{3,.7}\right) \alpha_{1,.7}^2 \Dtau \right)^{-\frac{1}{2}} \\
		\alpha_{2,.7} \exp \left(-\frac{1}{4 m_3}\alpha_{1,.7}^3 \pi_{1,.7} \Dtau \right) \\
		\alpha_{3,.7} \exp \left(-\frac{1}{4 m_2}\alpha_{1,.7}^3 \pi_{1,.7} \Dtau \right) \\
		\pi_{1,.7} \left(1 + \frac{1}{2} \left(\frac{1}{m_3} \alpha_{2,.7} \pi_{2,.7} + \frac{1}{m_2} \alpha_{3,.7} \pi_{3,.7}\right) \alpha_{1,.7}^2 \Dtau \right)^{\frac{3}{2}} \\
		\pi_{2,.7} \exp \left(\frac{1}{4 m_3}\alpha_{1,.7}^3 \pi_{1,.7} \Dtau \right) \\
		\pi_{3,.7} \exp \left(\frac{1}{4 m_2}\alpha_{1,.7}^3 \pi_{1,.7} \Dtau \right)
	\end{pmatrix} \\
\bb{\xi}_9 =&
	\begin{pmatrix}
		\alpha_{1,.8} \exp \left(-\frac{1}{4 m_3}\alpha_{2,.8}^3 \pi_{2,.8} \Dtau \right) \\
		\alpha_{2,.8} \left(1 + \frac{1}{2} \left(\frac{1}{m_1} \alpha_{3,.8} \pi_{3,.8} + \frac{1}{m_3} \alpha_{1,.8} \pi_{1,.8}\right) \alpha_{2,.8}^2 \Dtau \right)^{-\frac{1}{2}} \\
		\alpha_{3,.8} \exp \left(-\frac{1}{4 m_1}\alpha_{2,.8}^3 \pi_{2,.8} \Dtau \right) \\
		\pi_{1,.8} \exp \left(\frac{1}{4 m_3}\alpha_{2,.8}^3 \pi_{2,.8} \Dtau \right) \\
		\pi_{2,.8} \left(1 + \frac{1}{2} \left(\frac{1}{m_1} \alpha_{3,.8} \pi_{3,.8} + \frac{1}{m_3} \alpha_{1,.8} \pi_{1,.8}\right) \alpha_{2,.8}^2 \Dtau \right)^{\frac{3}{2}} \\
		\pi_{3,.8} \exp \left(\frac{1}{4 m_1}\alpha_{2,.8}^3 \pi_{2,.8} \Dtau \right)
	\end{pmatrix} \\
\bb{\xi}_{10} =&
	\begin{pmatrix}
		\alpha_{1,.9} \exp \left(-\frac{1}{4 m_2}\alpha_{3,.9}^3 \pi_{3,.9} \Dtau \right) \\
		\alpha_{2,.9} \exp \left(-\frac{1}{4 m_1}\alpha_{3,.9}^3 \pi_{3,.9} \Dtau \right) \\
		\alpha_{3,.9} \left(1 + \frac{1}{2} \left(\frac{1}{m_2} \alpha_{1,.9} \pi_{1,.9} + \frac{1}{m_1} \alpha_{2,.9} \pi_{2,.9}\right) \alpha_{3,.9}^2 \Dtau \right)^{-\frac{1}{2}} \\
		\pi_{1,.9} \exp \left(\frac{1}{4 m_2}\alpha_{3,.9}^3 \pi_{3,.9} \Dtau \right) \\
		\pi_{2,.9} \exp \left(\frac{1}{4 m_1}\alpha_{3,.9}^3 \pi_{3,.9} \Dtau \right) \\
		\pi_{3,.9} \left(1 + \frac{1}{2} \left(\frac{1}{m_2} \alpha_{1,.9} \pi_{1,.9} + \frac{1}{m_1} \alpha_{2,.9} \pi_{2,.9}\right) \alpha_{3,.9}^2 \Dtau \right)^{\frac{3}{2}}
	\end{pmatrix} = \bb{z}_1.
\end{align*}

\end{appendices}

\bibliographystyle{plain}
\bibliography{bibliography}

\end{document}